\documentclass[12pt]{article}
\usepackage[top=2.5cm, bottom=2.5cm, left=2.5cm, right=2.5cm]{geometry}
\usepackage{graphicx,amsmath,amssymb}
\usepackage{color}
\usepackage{subfigure}
\usepackage{appendix}
\usepackage{chemist} 
\usepackage[version=3]{mhchem}
\newcommand{\dd}{\textrm{d}}
\usepackage{epstopdf}
\usepackage{authblk}

\title{Elimination of fast variables in stochastic nonlinear kinetics}

\author[1,2]{Gabriel Morgado}
\author[1,3]{Bogdan Nowakowski}
\author[2]{Annie Lemarchand}
\affil[1]{Institute of Physical Chemistry, Polish Academy of Sciences, Kasprzaka 44/52, 01-224 Warsaw, Poland}
\affil[2]{Laboratoire de Physique Th\'eorique de la Mati\`ere Condens\'ee, Sorbonne Universit\'e, CNRS , 4 place Jussieu, case courrier 121, 75252 Paris CEDEX 05, France}
\affil[3]{SGGW, Warsaw University of Life Sciences, Nowoursynowska 159, 02-776 Warsaw, Poland}

\date{}
\begin{document}
\maketitle

\begin{abstract}
A reduced chemical scheme involving a small number of variables is often sufficient to account for the deterministic evolution of the concentrations of the main species contributing to a reaction. However its predictions are questionable in small systems used for example in fluorescence correlation spectroscopy (FCS) or in explosive systems involving strong nonlinearities such as autocatalytic steps.
We make precise dynamical criteria defining the validity domain of the quasi-steady-state approximation and the elimination of a fast concentration in deterministic dynamics.
Designing two different three-variable models converging toward the same two-variable model, we show that the variances and covariance of the fluctuations of the slow variables are not correctly predicted by the two-variable model, even in the limit of a large system size. The more striking weaknesses of the reduced scheme are figured out in mesoscaled systems containing a small number of molecules.
The results of two stochastic approaches are compared and the shortcomings of the Langevin equations with respect to the master equation are pointed out.
We conclude that the description of the fluctuations and their coupling with nonlinearities of deterministic dynamics escape reduced chemical schemes. 
\end{abstract}
\baselineskip=24pt

\section{Introduction}
The quasi-steady-state approximation is currently used in chemistry to eliminate a fast variable and build 
tractable reaction mechanisms involving a few species while satisfactorily accounting for the dynamical behavior
of the system \cite{vankampen,segel,turanyi,gorban}. 
However, in a small system, containing a small number of particles, fluctuations reach large amplitudes \cite{nicolis}. Our goal is to make precise how eliminating a fast variable in a model of growing Turing pattern \cite{turing} may affect the prediction on the amplitude of concentration fluctuations. The interplay between nonlinearities of deterministic dynamics and fluctuations makes the elimination of fast variables in stochastic dynamics non trivial \cite{moreau,papoian,sinitsyn,thomas,constable13,constable14}.
The issue is essential to estimate the ability of a reduced chemical scheme to report on experiments in small samples such as fluorescence correlation spectroscopy (FCS) experiments \cite {fcs,fcsludo}. 
To this purpose we consider a minimal chemical model involving two species of variable concentrations, able
to reproduce the propagation of a chemical wave front toward a stable steady state and a spatially-periodic structure of Turing type \cite{epl11,jcp139,pre16,pre18}. 
The minimum model is assumed to result from the reduction of three-variable models.
The question is to determine if the amplitude of the fluctuations deduced from a three-variable model is correctly predicted by the two-variable model, 
in the limit where the reduction of deterministic dynamics is valid and for homogeneous conditions in the vicinity of a steady state.\\

The paper is organized as follows. In section 2, we determine the conditions on the parameters for the deterministic dynamics of the two- and three-variable models to be as close as possible.
In particular, we look for conditions ensuring that the steady states coincide in order to make the comparison between the models relevant.
Section 3 is devoted to the stochastic description of the different models.
Langevin equations with internal noise \cite{gillange} deduced from the chemical master equation are used to derive analytical expressions of the variances and covariances of concentration fluctuations for the two-variable and three-variable models \cite{jcp140,jcp141,physica15}.
Choosing the nontrivial stable steady state of the two-variable and three-variable models as initial condition, we simulate the master equation
using the kinetic Monte Carlo method introduced by Gillespie \cite{gillespie} and determine the variances and covariance of the fluctuations of the slow variables. The accuracy of the Langevin approach is checked by comparing the results given by the two stochastic approaches.
Section 4 is devoted to discussion and conclusion.

\section{Deterministic dynamics}
\subsection{The two-variable model}
The following chemical mechanism, inspired by the Schnakenberg model \cite{schnakenberg} and the Gray-Scott model \cite{gray}, has been designed 
to account for different self-organization phenomena in open systems \cite{murray}
\begin{eqnarray}
\label{reac1II}
\ce{X ->[k_1] R_1}\\
\label{reac2II}
\ce{2X + Y ->[k_2] 3X}\\
\label{reac3II}
\ce{Y <=>[k_3][k'_{-3}] R_2}
\end{eqnarray}
In particular, we used it to study the impact of fluctuations \cite{epl11} on growing Turing structures with possible application to 
the development of periodic patterns in embryos \cite{jcp139,pre16,pre18}.
The concentrations of the species R$_1$ and R$_2$ are assumed to be constant through appropriate matter exchanges with reservoirs, 
also called chemostats by analogy with thermostats capable of fixing temperature through heat exchanges.
The model involves two species X and Y of variable concentrations $X$ and $Y$ governed by the following differential equations
\begin{eqnarray}
\label{eqXII}
\frac{\dd X}{\dd t} &=& -k_1X+k_2 X^2Y \\
\label{eqYII}
\frac{\dd Y}{\dd t} &=& k_{-3} - k_3Y - k_2 X^2Y
\end{eqnarray}
where the $k_i$'s, for $i=1,2,3,-3$, are rate constants. For the sake of simplicity, we set $k_{-3}=k'_{-3}R_2$.

For parameter values obeying $\Delta<0$ with $\Delta=k_{-3}^2-4k_1^2k_3/k_2$, the two-variable model admits a single steady state $(X_2^0=0, Y_2^0=k_{-3}/k_3)$.
If $\Delta \ge 0$, the model has three stationary states $(X_2^0,Y_2^0)$, $(X_1^0,Y_1^0)$, and $(X_0^0,Y_0^0)$ with 
\begin{eqnarray}
X_1^0&=&\frac{k_{-3}-\sqrt{\Delta}}{2k_1} \\
Y_1^0&=&\frac{k_{-3}+\sqrt{\Delta}}{2k_3} \\
\label{X0II}
X_0^0&=&\frac{k_{-3}+\sqrt{\Delta}}{2k_1} \\
\label{Y0II}
Y_0^0&=&\frac{k_{-3}-\sqrt{\Delta}}{2k_3} 
\end{eqnarray}

In the domain of existence of the three stationary states, $(X_1^0,Y_1^0)$ is unstable whereas $(X_0^0,Y_0^0)$ and $(X_2^0,Y_2^0)$ are stable.
The model has been designed to have a steady state obeying $X_2^0=0$, unable to create X species {\it ex nihilo} and consequently insensitive to internal fluctuations. 
Hence, in a spatially extended system, it is possible to prepare 
a region of space in the state $(X_2^0,Y_2^0)$ and study how it is invaded by a propagating chemical wave front. After the passage of the wave front, the system relaxes towards the steady state $(X_0^0,Y_0^0)$ and can be destabilized by inhomogeneous perturbations, being then replaced by a periodic spatial pattern of Turing type \cite{jcp139,pre16,pre18}.

In the following, we focus on a homogeneous system and study the linear dynamics in the vicinity of the steady state $(X_0^0,Y_0^0)$. 
Introducing the deviations $x=X-X_0^0$ and $y=Y-Y_0^0$ from the steady state, we locally characterize dynamics by 
the linearized equations
\begin{equation}
\label{linear}
\frac{{\rm d}\zeta}{{\rm d}t}={\mathbf M}\zeta
\end{equation}
where $\zeta=\begin{pmatrix}x \\ y\end{pmatrix}$ is the vector representing the deviation from the steady state and ${\mathbf M}$ is the stability matrix, given by
\begin{eqnarray}
\label{M}
{\mathbf M}=
\begin{pmatrix}
m_{11}=k_1   & m_{12}=k_2(X_0^0)^2 \\
m_{21}=-2k_1 & m_{22}=-k_3-k_2(X_0^0)^2 
\end{pmatrix}
\end{eqnarray}
The eigenvalues of ${\mathbf M}$ are
\begin{eqnarray}
\label{lambda1}
\lambda_1^0&=&\frac{k_1-k_3-k_2(X_0^0)^2+\sqrt{\Delta'}}{2} \\
\label{lambda2}
\lambda_2^0&=&\frac{k_1-k_3-k_2(X_0^0)^2-\sqrt{\Delta'}}{2}
\end{eqnarray}
with $\Delta'=(k_2(X_0^0)^2+k_3-k_1)^2-4k_1(k_2(X_0^0)^2-k_3)$ 
and $X_0^0$ given in Eq. (\ref{X0II}).
The vector $\chi=\begin{pmatrix}x_1 \\ x_2\end{pmatrix}$ of coordinates in the eigenbasis of ${\mathbf M}$ is related to the vector $\zeta$ through $\zeta={\mathbf P^0}\chi$ where
the change of basis matrix is
\begin{eqnarray}
\label{P}
{\mathbf P^0}=
\begin{pmatrix}
p_{11}^0=k_2(X_0^0)^2 & p_{12}^0=k_2(X_0^0)^2 \\
p_{21}^0=\lambda_1-k_1   & p_{22}^0=\lambda_2-k_1
\end{pmatrix}
\end{eqnarray}
The dynamics around the steady state is locally characterized by the two relaxation times $\tau_i^0=1/\mid R(\lambda_i^0) \mid$, with $i=1,2$, where $R$ returns the real part of the argument.

Two different three-variable models converging to the two-variable model after elimination of a fast variable $Z$ are introduced in the next subsections.

\subsection{Three-variable model $A$}
One of the simplest way to introduce a third variable concentration $Z$ consists in
considering an intermediate species Z \cite{cook}, reversibly formed through the second reaction with rate constants $k'_2$ and $k'_{-2}$ and irreversibly
transformed into 3X with rate constant $k''_2$ according to the three-variable model $A$ 
\begin{eqnarray}
\label{reac1IIIa}
\ce{X ->[k_1] R_1}\\
\label{reac2IIIa}
\ce{2X + Y <=>[k'_2][k'_{-2}] Z}\\
\label{reac3IIIa}
\ce{Z ->[k''_2] 3X}\\
\label{reac4IIIa}
\ce{Y <=>[k_3][k'_{-3}] R_2}
\end{eqnarray}
The goal of this subsection is to determine the conditions for which the three-variable model reduces
to the two-variable model.
The rate equations associated with model $A$ are
\begin{eqnarray}
\label{eqXIIIa}
\frac{\dd X}{\dd t} &=& -k_1X-2k'_2 X^2Y+(2k'_{-2}+3k''_2)Z\\
\label{eqYIIIa}
\frac{\dd Y}{\dd t} &=& k_{-3} - k_3Y - k'_2 X^2Y+k'_{-2}Z\\
\label{eqZIIIa}
\frac{\dd Z}{\dd t} &=& k'_2 X^2Y-(k'_{-2}+k''_2)Z
\end{eqnarray}
On the slow manifold \cite{fenichel,fraser,gorban} defined by $\frac{\dd Z}{\dd t}=0$, i.e. 
\begin{equation}
\label{sma}
Z=\frac{k'_2}{k'_{-2}+k''_2}X^2Y
\end{equation}
dynamics becomes
\begin{eqnarray}
\label{eqXIIIared}
\frac{\dd X}{\dd t} &=& -k_1X+\frac{k'_2k''_2}{k'_{-2}+k''_2} X^2Y\\
\label{eqYIIIared}
\frac{\dd Y}{\dd t} &=& k_{-3} - k_3Y - \frac{k'_2k''_2}{k'_{-2}+k''_2}X^2Y
\end{eqnarray}
which exactly matches the dynamics of the two-variable model given in Eqs. (\ref{eqXII},\ref{eqYII}) provided that:
\begin{equation}
\label{conda}
\frac{k'_2k''_2}{k'_{-2}+k''_2}=k_2
\end{equation}
In particular, when the above condition is satisfied, the steady state $(X_0,Y_0,Z_0)$ of model $A$ obeys
\begin{eqnarray}
\label{X0IIIa}
X_0&=&X_0^0\\
\label{Y0IIIa}
Y_0&=&Y_0^0\\
\label{Z0IIIa}
 Z_0&=&\frac{k_1X_0^0}{k''_2}
\end{eqnarray}
where the expression of the steady concentrations $X_0^0$ and $Y_0^0$ for the two-variable model are given in Eqs. (\ref{X0II},\ref{Y0II}).
Provided that Eq. (\ref{conda}) is obeyed, the steady concentrations of the three-variable model $A$ do not depend on the rate constants $k'_2$ and $k'_{-2}$.
The variation of the steady concentrations versus the rate constant $k''_2$ is given in Fig. 1 
for the two-variable model and the three-variable model $A$. 
For the chosen parameter values, the steady concentration $Z_0$ of the eliminated species is never negligible with respect to $X_0^0$ and $Y_0^0$. However this qualitative statement cannot be considered as a criterion to check the validity of the quasi-steady-state approximation, which refers to dynamics and not to steady properties.
\begin{figure}
\centering
\hspace{-2cm}
\includegraphics[scale=0.6]{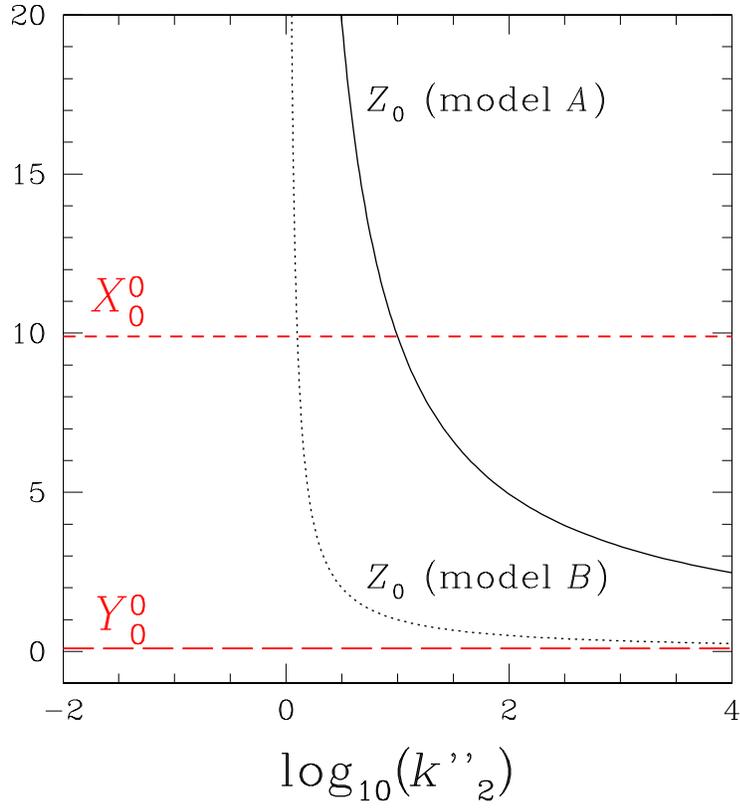}
\caption{Steady concentrations of species X and Y for the two-variable model and the three-variable models $A$ and $B$ (red short-dashed line for $X_0=X_0^0$ and red long-dashed line for $Y_0=Y_0^0$) and steady concentration $Z_0$ of the three-variable model $A$ (black solid line) and the three-variable model $B$ (black dotted line) versus $\log_{10}(k''_2)$ for $k_1=k_2=k_3=1$, $k_{-3}=10$, and $\frac{k'_2k''_2}{k'_{-2}+k''_2}=k_2$. 
}
%1
\end{figure}

The dynamics of the three-variable model depends on the three rate constants $k'_2$, $k'_{-2}$, and $k''_{2}$, in addition to the parameters of the two-variable model. We study the behavior of the system in two cuts of the parameter space $(k'_2, k'_{-2}, k''_2)$. 
In case $a$, the parameter $k'_{-2}$ is set at $k'_{-2}=1$ and $k'_2$ varies with $k''_2$ according to $k'_2=k_2(1+k'_{-2}/k''_2)$ 
in order to obey Eq. (\ref{conda}). 
Case $b$ corresponds to $k'_2=10$ and $k'_{-2}=k''_2(-1+k'_2/k_2)$, which also obeys Eq. (\ref{conda}). 
We adopt analogous notations as in the two-variable model to characterize the linearized dynamics
around the steady state. In particular, the deviations from the steady state are denoted by $x=X-X_0^0$, $y=Y-Y_0^0$, and $z=Z-Z_0$.
The matrix form of the linearized equations around the steady state are similar to Eq. (\ref{linear}) with
$\zeta=\begin{pmatrix}x \\ y \\ z\end{pmatrix}$ and a $3\times 3$ matrix ${\mathbf M}$ associated with eigenvalues $\lambda_i$, with $i=1,2,3$.
The vectors $\chi=\begin{pmatrix}x_1 \\ x_2 \\x_3\end{pmatrix}$ and $\zeta$ are related by $\zeta={\mathbf P}\chi$,
where ${\mathbf P}$ is the $3\times 3$ change of basis matrix. The expressions of the eigenvalues and eigenvectors associated with the three-variable model $A$ are given in Appendix A.
The variation of the eigenvalues $\lambda_i$, for $i=1,2,3$, versus $k''_2$ is given in Fig. 2. The eigenvalues $\lambda_1^0$ and $\lambda_1$ of the two- and three-variable models are identical in the entire $k''_2$ range.
\begin{figure}
\centering
\hspace{-2cm}
\subfigure{\includegraphics[scale=0.4]{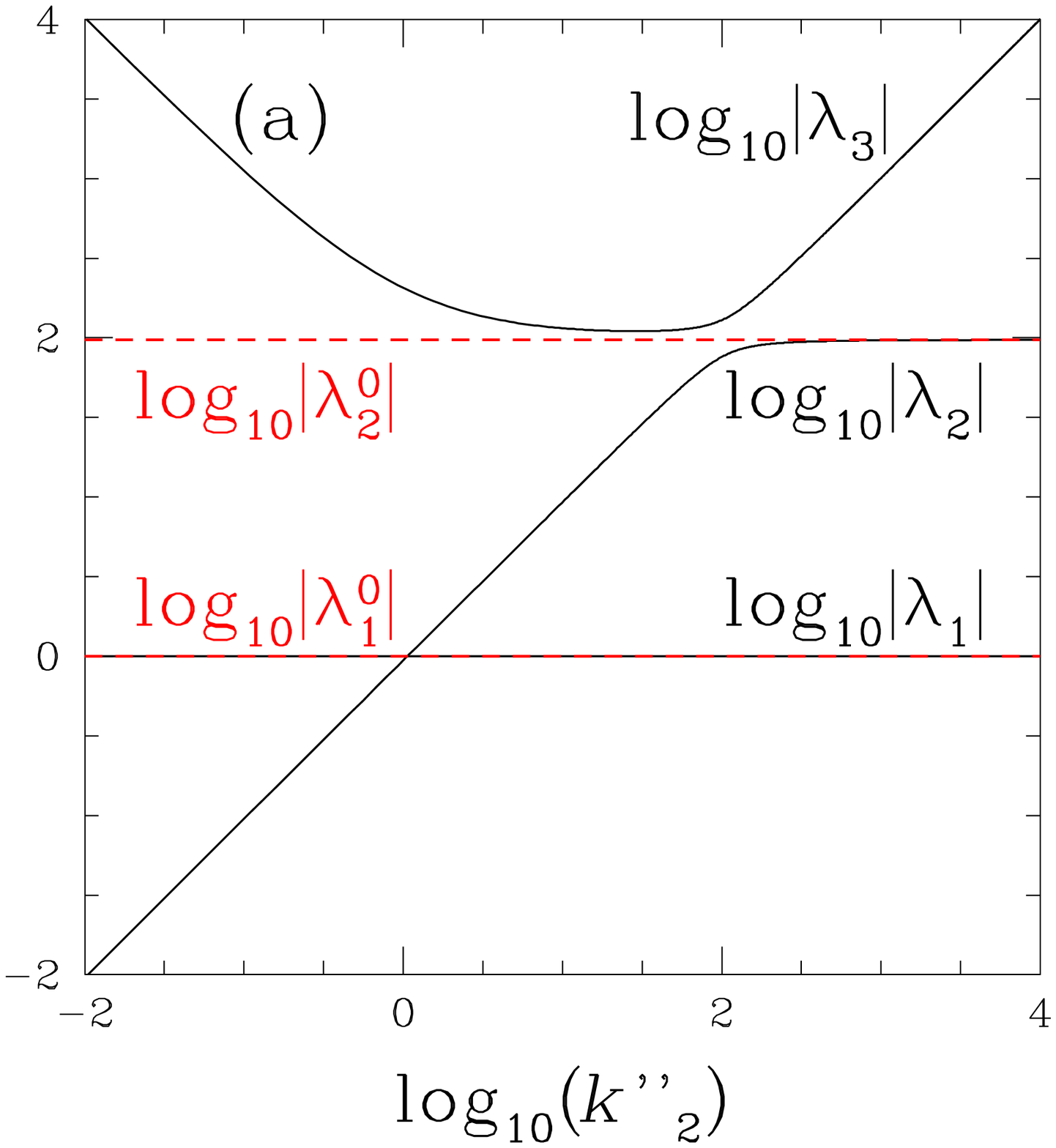}}
\subfigure{\includegraphics[scale=0.4]{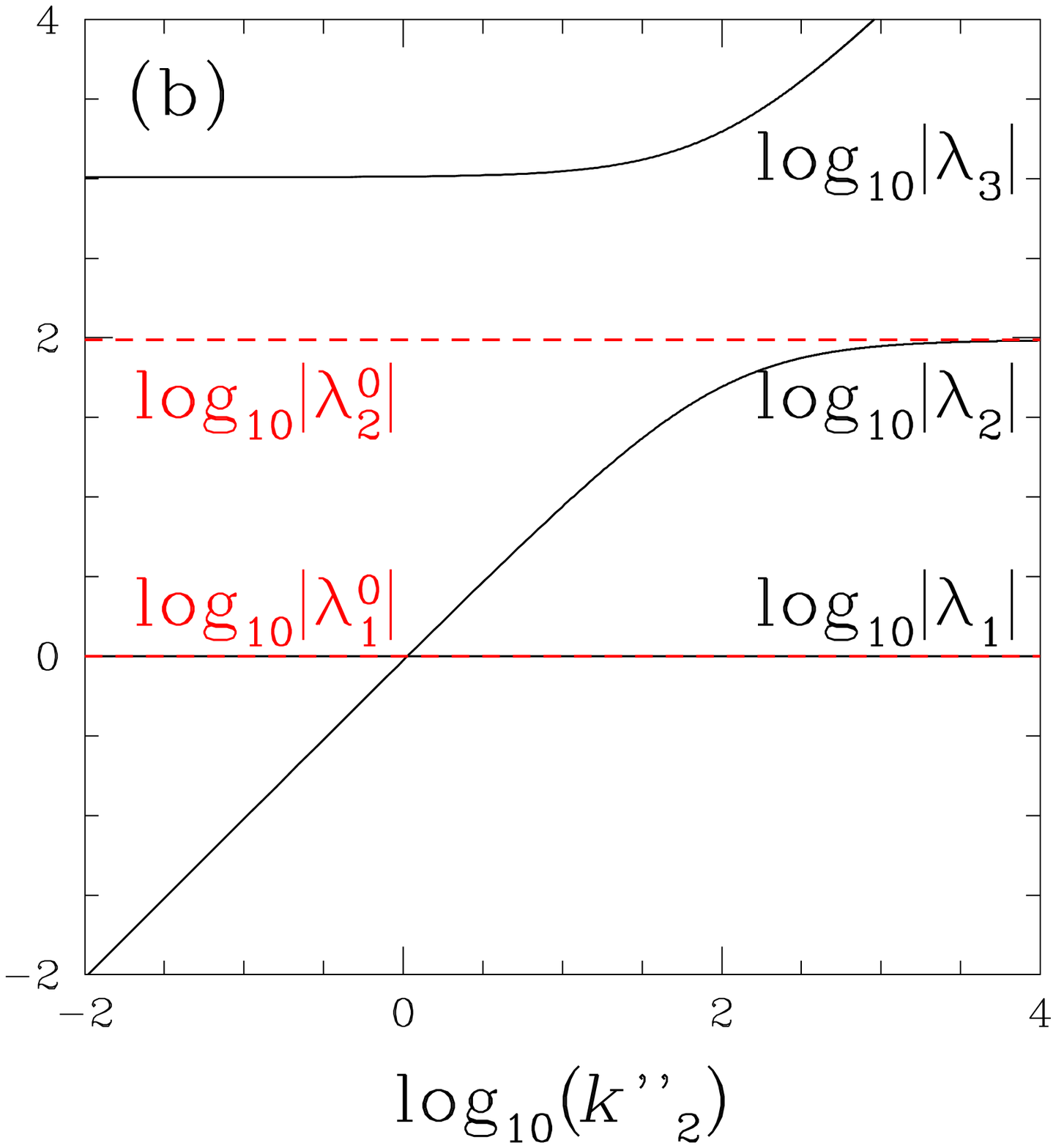}}
\caption{Logarithm of absolute eigenvalues $\log_{10}\mid \lambda_i^0 \mid$, $i=1,2$, of the two-variable model (red dashed lines) and $\log_{10}\mid \lambda_i \mid$, $i=1,2,3$, of the three-variable model $A$ (black solid lines) versus $\log_{10}(k''_2)$ for $k_1=k_2=k_3=1$, $k_{-3}=10$. All eigenvalues are real.}
(a) Case $a$: $k'_{-2}=1$ with $k'_2=k_2(1+k'_{-2}/k''_2)$.
(b) Case $b$: $k'_2=10$ with $k'_{-2}=k''_2(-1+k'_2/k_2)$.
%2
\end{figure}
The time $\tau_i=1/\mid R(\lambda_i) \mid$ associated with the variable $x_i$ characterizes the evolution along the corresponding eigendirection.

In the linear domain around the steady state, the
elimination of the variable $Z$ according to the quasi-steady-state approximation can be performed if two conditions are fulfilled. First, the
relaxation in one eigendirection must be particularly fast. This condition implies that the
real part of the eigenvalue associated with the fast eigendirection, for example $\lambda_3$, must be significantly larger in absolute value than the two others \cite{segel,turanyi,gorban}:
\begin{equation}
\label{condlam}
\mid R(\lambda_3) \mid > \max(\mid R(\lambda_1) \mid, \mid R(\lambda_2) \mid)
\end{equation}
Second, the coordinate $z$ must vary in the same way as $x_3$ during the time interval $[0,\tau_3]$. This condition can be expressed using the inverse change of basis matrix ${\mathbf P^{-1}}$
\begin{equation}
\label{condinv}
x_3=q_{31}x+q_{32}y+q_{33}z
\end{equation}
leading to
\begin{equation}
\label{condvecpr}
\mid R(q_{33}) \mid > \max(\mid R(q_{31}) \mid, \mid R(q_{32}) \mid)
\end{equation}
where $q_{3i}$, for $i=1,2,3$, are the elements of the third line of ${\mathbf P^{-1}}$ and for initial conditions with departures from the steady states $x$, $y$, and $z$ of the same order of magnitude.
An equivalent condition can be written using the change of basis matrix ${\mathbf P}$, provided that its column vectors, i.e. the eigenvectors, are normalized
\begin{equation}
\label{condir}
z=p_{31}x_1+p_{32}x_2+p_{33}x_3
\end{equation}
which reads
\begin{equation}
\label{condvecpr0}
\mid R(p_{33}) \mid > \max(\mid R(p_{31}) \mid, \mid R(p_{32}) \mid)
\end{equation}
where $p_{3i}$, for $i=1,2,3$, are the elements of the third line of ${\mathbf P}$.
Hence, the eigendirections associated with the two small absolute real parts of the eigenvalues are close to the plan $z=0$.
If the two conditions given in Eqs. (\ref{condlam}) and (\ref{condvecpr}) are fulfilled, the relaxation with the short characteristic time $\tau_3$ along the $x_3$-axis can be considered as instantaneous. Further evolution, including $Z$ evolution, occurs with the slower relaxation times on the slow manifold given in Eq. (\ref{sma}). The slow manifold is tangent to the slow eigendirections and is close to the 
$z=0$ plane in the vicinity of the steady state.

\begin{figure}
\hspace{-2cm}
\centering
\subfigure{\includegraphics[scale=0.4]{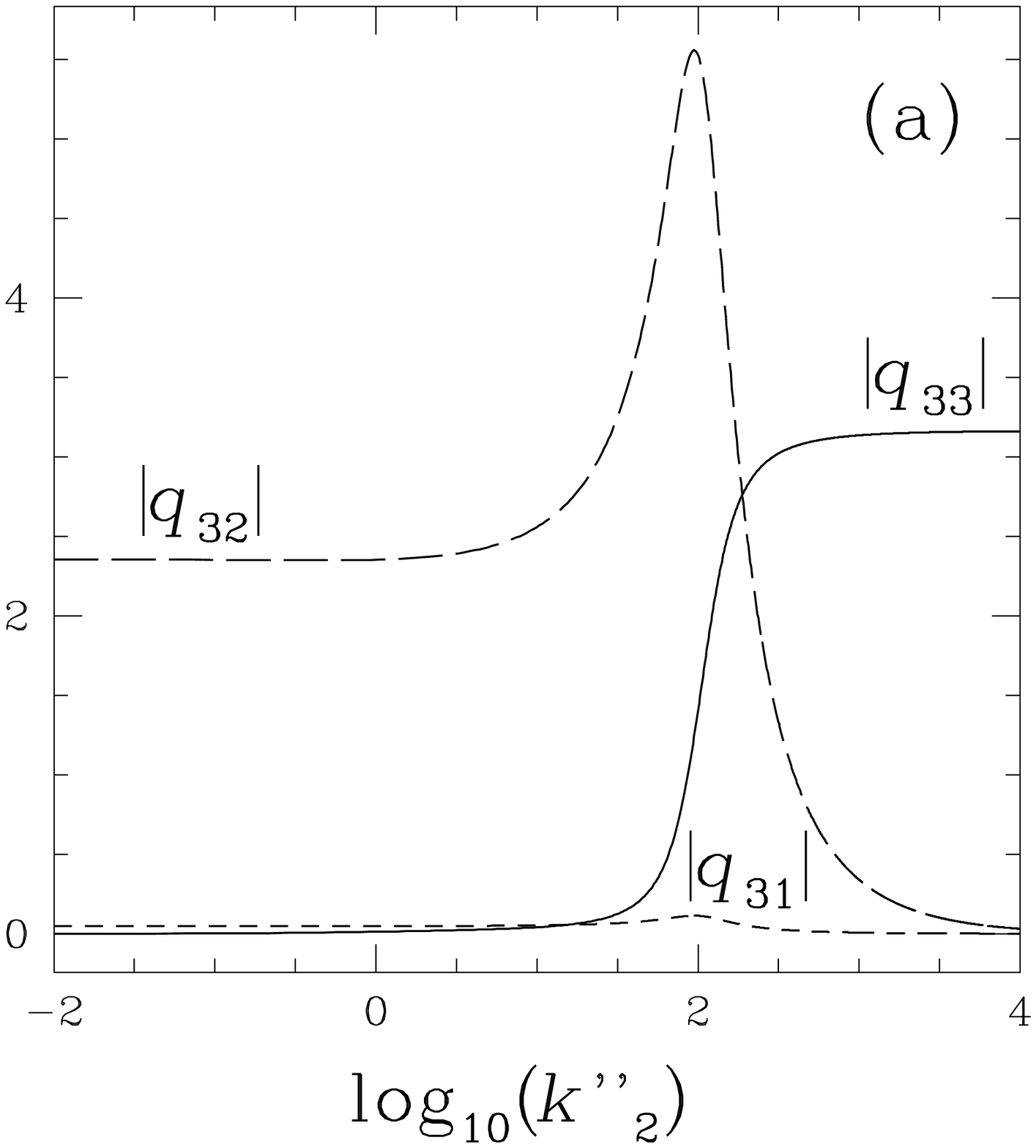}}
\subfigure{\includegraphics[scale=0.4]{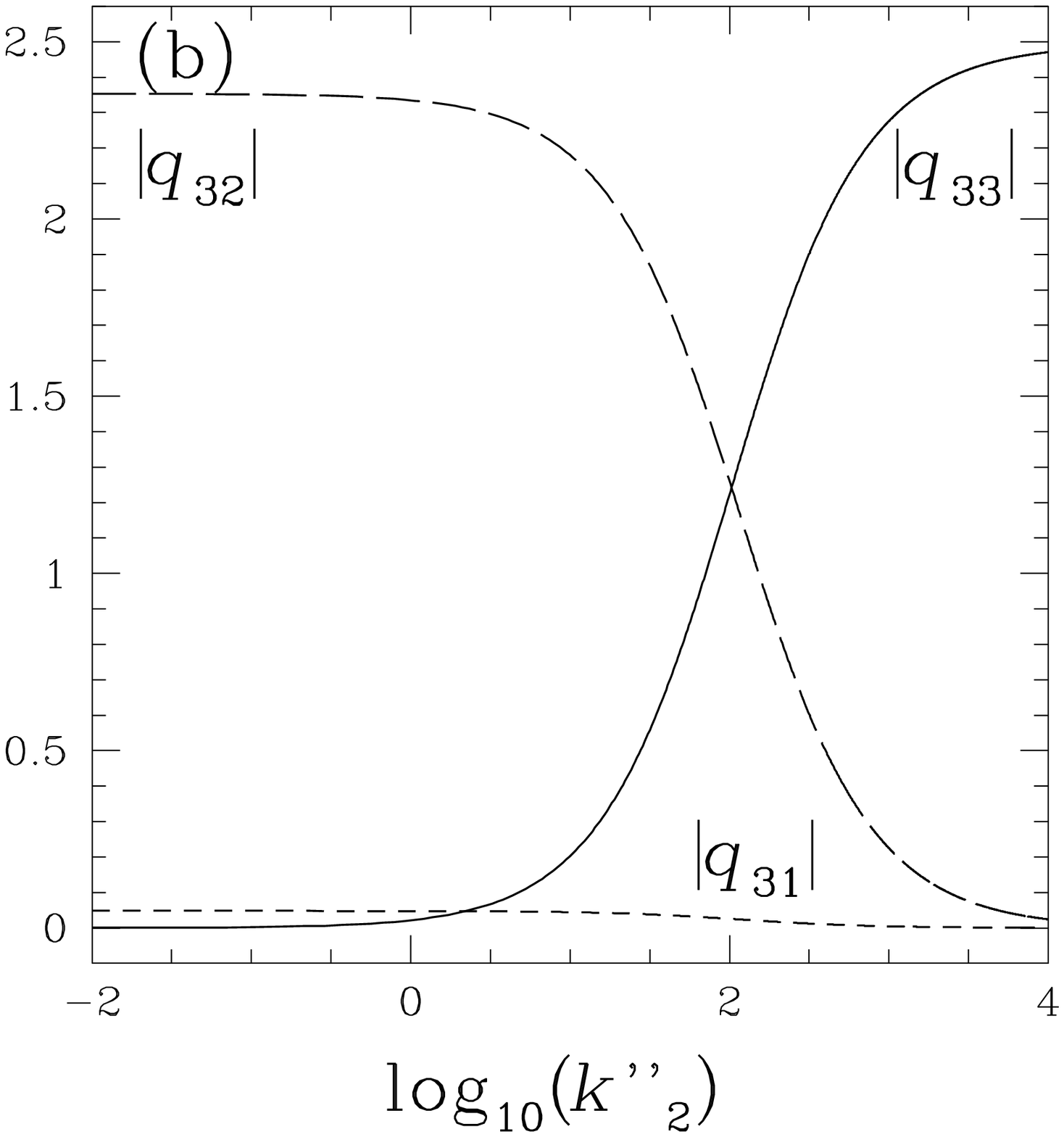}}
\caption{Three-variable model $A$: Absolute elements $\mid q_{31}\mid$ (short-dashed line), $\mid q_{32}\mid$ (long-dashed lines), and $\mid q_{33}\mid$ (solid line) of the inverse change of basis matrix versus rate constant $\log_{10}(k''_2)$ for the parameter values given in the caption of Fig. 2. (a) Case $a$ and (b) Case $b$.}
%3
\end{figure}

According to Fig. 2 and for the chosen parameter values, the eigenvalues of the three-variable model $A$ are always real, negative, and obey Eq. (\ref{condlam}) except for $k''_2$ around $100$ in case $a$. The variable $x_3$ of model $A$ which evolves with the characteristic time $\tau_3=1/\mid R(\lambda_3) \mid$ can be considered as fast with respect to the variables $x_1$ and $x_2$ nearly in the entire range of $k''_2$ values. The eigenvalues $\lambda_1$ and $\lambda_2$ of the three-variable model $A$ coincide with the eigenvalues $\lambda_1^0$ and $\lambda_2^0$ of the two-variable system in a smaller interval, $k''_2 > 100$.\\

Using the vocabulary of quantum chemistry, we notice that an avoided crossing between $\lambda_2$ and $\lambda_3$ is observed in Fig. 2a for $k''_2 \simeq 100$. The parallelism with a known case of failure of the Born-Oppenheimer approximation can be drawn. Both the quasi-steady-state approximation and the Born- Oppenheimer approximation belong to the class of adiabatic approximations \cite{vankampen}. The Born-Oppenheimer approximation consists in ignoring the fast movements of the electrons and only considering the slow components of their displacements in response to the movement of the nuclei. Introducing a perturbation to the Hamiltonian splits the degenerate energy states and leads to an avoided crossing, exactly as in Fig. 2a when switching from the two-variable model to the three-variable model.
The cut of the parameter space associated with case $b$ does not lead to an avoided crossing for $\lambda_2$ and $\lambda_3$ as shown in Fig. 2b.
It is to be noted that, for $k''_2 =1$, the crossing of the two eigenvalues $\lambda_1$ and $\lambda_2$ associated with the slow dynamics is not avoided in model $A$ in both cases $a$ and $b$.\\

As shown in Fig. 3, imposing that $z$ evolves like $x_3$ is more restrictive than Eq. (\ref{condlam}). The condition on the elements of ${\mathbf P^{-1}}$ given in Eq. (\ref{condvecpr}) is obeyed for $k''_2 > 10^{2.48}$ in cases $a$ and $b$. Eq. (\ref{condvecpr0}) based on the elements of ${\mathbf P}$ leads to similar results. The avoided crossing of $\lambda_3$ and $\lambda_2$ observed in Fig. 2a is associated with a maximum for $\mid R(q_{32}) \mid$ in Fig. 3a. We note the absence of both phenomena in Figs. 2b and 3b.

We conclude that the two conditions given in Eqs. (\ref{condlam}) and (\ref{condvecpr}) are satisfied in the range $k''_2 > 10^{2.48}$ in which the linear dynamics of the two-variable model and the three-variable model $A$ are close for the two cases $a$ and $b$. It is worth noting that Eqs. (\ref{condlam}) and (\ref{condvecpr}) are only necessary conditions
of validity of the quasi-steady-state approximation. But they do not warrant that the two-variable model remains valid when nonlinearities become important, 
typically when the system is far from the steady state or close to a bifurcation.

\subsection{Three-variable model $B$}
The second reaction of the two-variable model can also be decomposed into two reactions involving an intermediate species Z according to
\begin{eqnarray}
\label{reac1III}
\ce{X ->[k_1] R_1}\\
\label{reac2III}
\ce{X + Y <=>[k'_2][k'_{-2}] Z}\\
\label{reac3III}
\ce{X + Z ->[k''_2] 3X}\\
\label{reac4II}
\ce{Y <=>[k_3][k'_{-3}] R_2}
\end{eqnarray}
leading to the following differential equations:
\begin{eqnarray}
\label{eqXIIIb}
\frac{\dd X}{\dd t} &=& -k_1X-k'_2 XY+k'_{-2}Z+2k''_2XZ\\
\label{eqYIIIb}
\frac{\dd Y}{\dd t} &=& k_{-3} - k_3Y - k'_2 XY+k'_{-2}Z\\
\label{eqZIIIb}
\frac{\dd Z}{\dd t} &=& k'_2 XY-k'_{-2}Z-k''_2XZ
\end{eqnarray}
The slow manifold $\frac{\dd Z}{\dd t}=0$ is now given by $Z=\frac{k'_2 XY}{k'_{-2}+k''_2X}$.
The dynamics on the slow manifold obeys
\begin{eqnarray}
\label{eqXIIIbred}
\frac{\dd X}{\dd t} &=& -k_1X+\frac{k'_2k''_2}{k'_{-2}+k''_2X} X^2Y\\
\label{eqYIIIbred}
\frac{\dd Y}{\dd t} &=& k_{-3} - k_3Y - \frac{k'_2k''_2}{k'_{-2}+k''_2X}X^2Y
\end{eqnarray}
which never rigorously converges to the dynamics associated with the two-variable model given in Eqs. (\ref{eqXII}) and (\ref{eqYII}).
Applying the quasi-steady-state approximation to model $B$ does make it possible to reduce the number of variables. 
However, due to the nonpolynomial form of Eqs. (\ref{eqXIIIbred}) and (\ref{eqYIIIbred}), the reduced equations cannot be interpreted as rate laws associated with a chemical mechanism involving elementary steps. In particular, the two-variable model defined by Eqs. (\ref{reac1II}-\ref{reac3II}) does not account for the nonlinearities of Eqs. (\ref{eqXIIIbred}) and (\ref{eqYIIIbred}) of order higher than $3$. 

Nevertheless, imposing the relation
\begin{equation}
\label{condb}
\frac{k'_2k''_2}{k'_{-2}+k''_2X_0^0}=k_2
\end{equation}
ensures that a steady state of model $B$ obeys
\begin{eqnarray}
\label{X0IIIb}
X_0&=&X_0^0\\
\label{Y0IIIb}
Y_0&=&Y_0^0\\
\label{Z0IIIb}
Z_0&=&k_1/k''_2
\end{eqnarray}
where $(X_0^0,Y_0^0)$ is the steady state of interest of the two-variable model. 
The variation of the steady concentrations of model $B$ versus the rate constant $k''_2$ are given in Fig. 1. As $k''_2$ increases, the value of $Z_0$ decreases earlier in the case of model $B$ than model $A$, which cannot be considered at this stage as indicating a larger domain of validity of the quasi-steady-state approximation applied to model $B$.

\begin{figure}
\centering
\hspace{-2cm}
\subfigure{\includegraphics[scale=0.4]{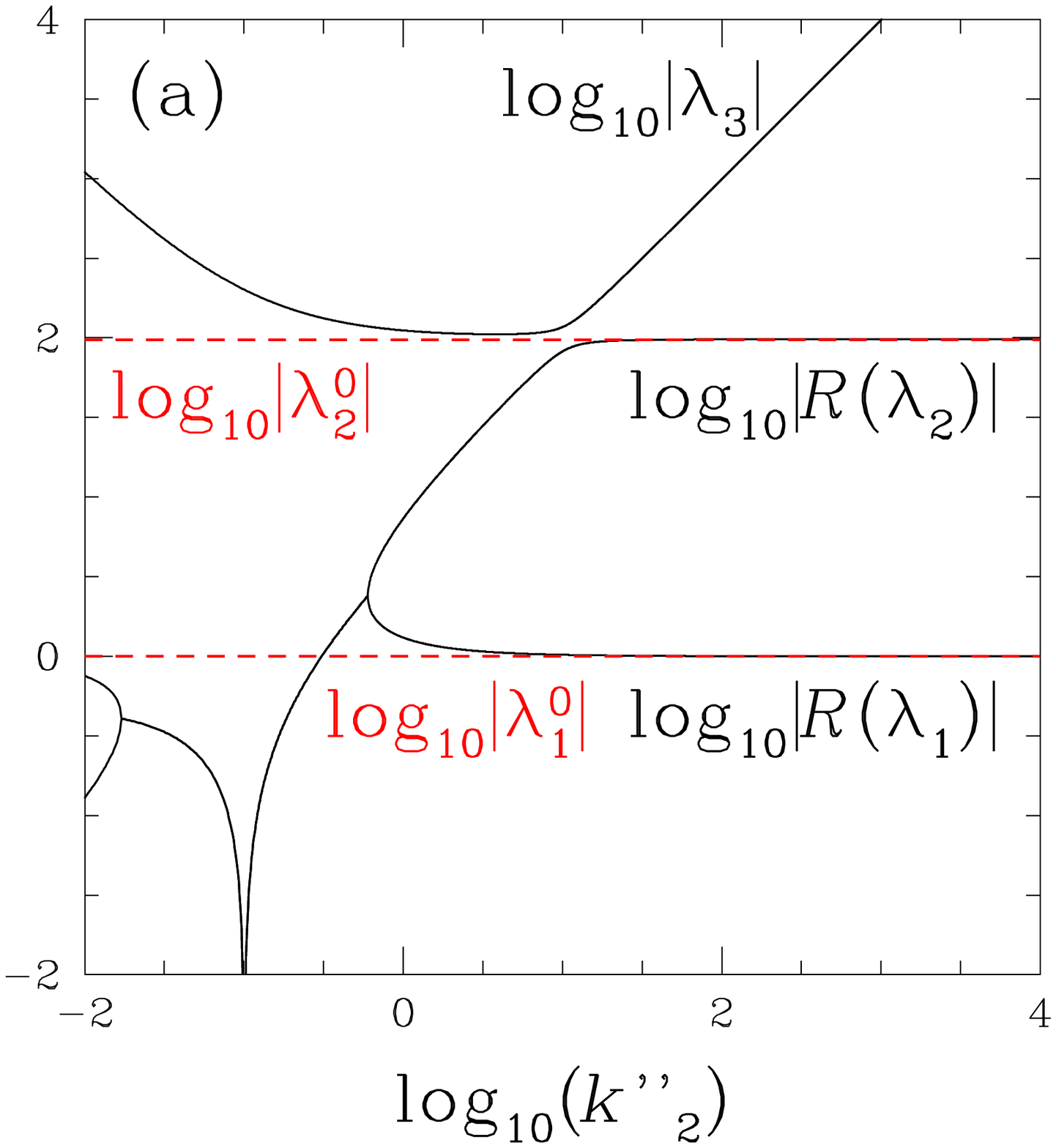}}
\subfigure{\includegraphics[scale=0.4]{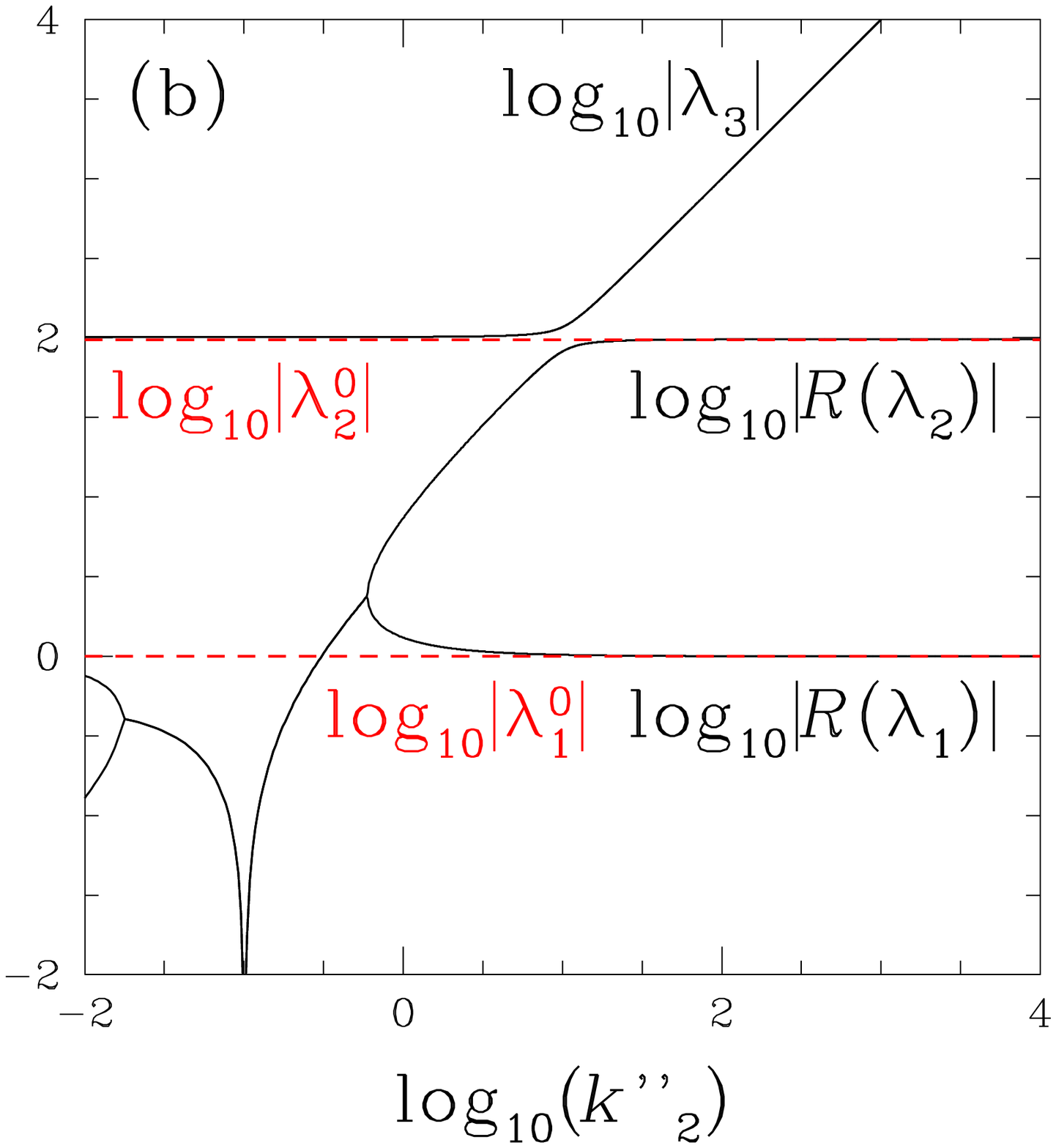}}
\caption{Same caption as in Fig. 2 for the three-variable model $B$ using the real part $R(\lambda_i)$ ($i=1,2$) of the eigenvalues when they are complex. 
(a) Case $a$: $k'_{-2}=1$ with $k'_2=k_2(X_0^0+k'_{-2}/k''_2)$.
(b) Case $b$: $k'_2=10$ with $k'_{-2}=k''_2(-X_0^0+k'_2/k_2)$.}
%4
\end{figure}

As shown in Fig. 4ab, the variation of the eigenvalues of the three-variable model $B$ with $k''_2$ are very similar in the two cases $a$ and $b$. As $k''_2$ decreases, several bifurcations are observed.
Three real eigenvalues are observed in the range $k''_2>10^{-0.22}$. 
As for model $A$, the two eigenvalues $\lambda_1$ and $\lambda_2$ coalesce for $k''_2=10^{-0.22}$. Only two real parts of eigenvalues are observed in the range 
$10^{-1.75}<k''_2<10^{-0.22}$ revealing the existence of two complex conjugate eigenvalues $\lambda_1$ and $\lambda_2$. 
As $k''_2$ becomes smaller than $0.1$, the real part of $\lambda_1$ and $\lambda_2$ becomes positive, as evidenced by the divergence of the logarithm of $\mid R(\lambda_1) \mid=\mid R(\lambda_2) \mid$ observed for $k''_2=0.1$. The steady state $(X_0^0,Y_0^0,Z_0)$ is then unstable. 
Another bifurcation occurs for $k''_2 \simeq 10^{-1.75}$, for which the eigenvalues become real again but remain negative.
If one excepts the neighborhood of $k''_2=10$ for which the orders of magnitude of $\lambda_2$ and $\lambda_3$ are comparable, model $B$ obeys the condition given in Eq. (\ref{condlam}) in the entire range of explored $k''_2$ values.
The two eigenvalues $\lambda_1^0$ and $\lambda_2^0$ of the two-variable model coincide with the eigenvalues $\lambda_1$ and $\lambda_2$ of the three-variable model $B$ if $k''_2 > 10$ in the two cases $a$ and $b$.

\begin{figure}
\centering
\subfigure{\includegraphics[scale=0.4]{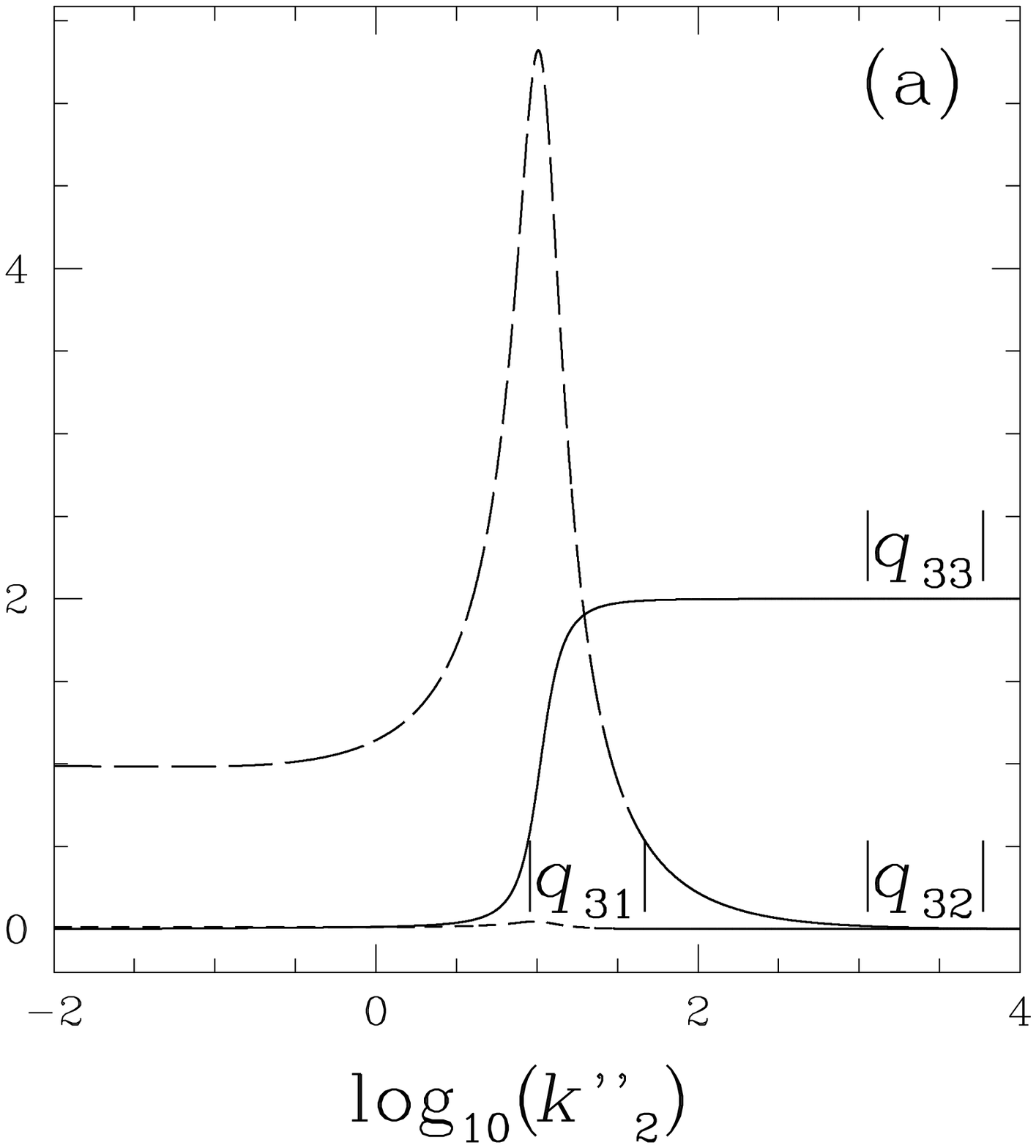}}
\subfigure{\includegraphics[scale=0.4]{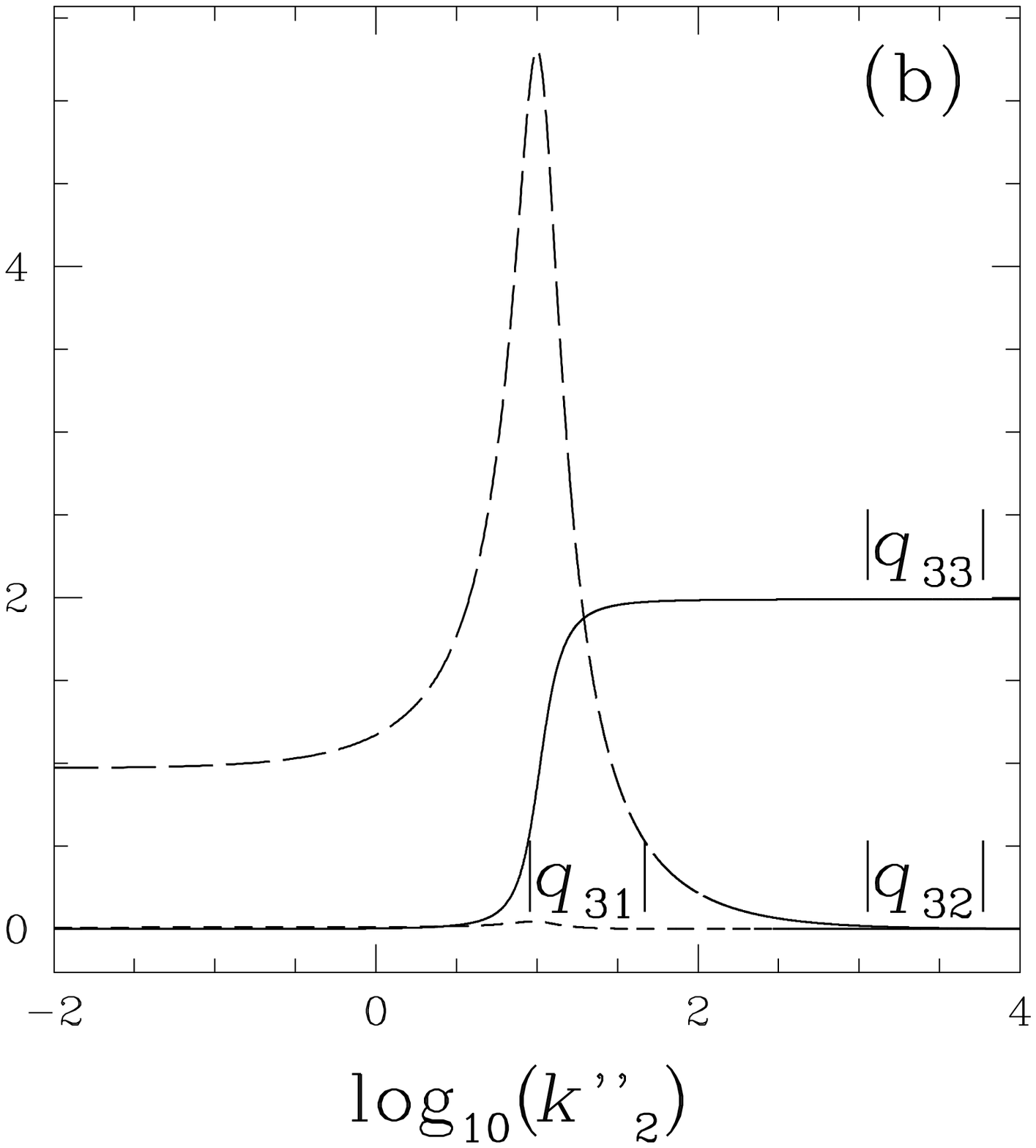}}
\caption{Same caption as in Fig. 3 for the three-variable model $B$. 
(a) Case $a$: $k'_{-2}=1$ with $k'_2=k_2(X_0^0+k'_{-2}/k''_2)$.
(b) Case $b$: $k'_2=10$ with $k'_{-2}=k''_2(-X_0^0+k'_2/k_2)$.
}
%5
\end{figure}
According to Fig. 5, the variation of the elements of the inverse change of basis matrix versus $k''_2$ are similar for the two cases $a$ and $b$. The avoided crossing between $\lambda_2$ and 
$\lambda_3$ observed in Fig. 4ab reveals that $x_2$ and $x_3$ evolve with analogous characteristic times and Eq. (\ref{condlam}) is not obeyed for $k''_2 \simeq 10$. This phenomenon is accompanied by a high peak for $\mid R(q_{32}) \mid$ and a significant departure from the condition given in Eq. (\ref{condvecpr}). The condition imposing that the variable $z$ behaves like the fast variable $x_3$ is obeyed in the range $k''_2>10^{1.48}$. Following the linear analysis, we conclude that the quasi-steady-state approximation is valid in the range $k''_2>10^{1.48}$ for the three-variable model $B$ in the two cases $a$ and $b$.

\section{Stochastic descriptions}
The effect of internal noise in a chemical system can be modeled in an approximate way within the framework of Langevin equations
using the expressions of the Langevin forces deduced from the chemical master equation \cite{gardiner,gillespie}.
In this section, we consider Langevin equations linearized in the vicinity of the steady state to compute approximate analytical expressions of variances and covariance of fluctuations for the slow concentrations $X$ and $Y$.
The method is illustrated in the case of the two-variable model. The Langevin equations are integrated in the eigenbasis. The inverse change of basis and the large-time limit are then used to find the steady variances of the concentrations \cite{jcp140,jcp141,physica15}.  
Deriving the corresponding expressions for the three-variable models is more tedious but follows the same intuitive procedure. The results are given in Appendix B. Another method, using an implicit matrix equation for the variances, is proposed by Gardiner \cite{gardiner} and leads to analogous results.\\

The linearized Langevin equations for the two-variable model are written as
\begin{eqnarray}
\frac{{\rm d}x}{{\rm d}t}&=& m_{11}x + m_{12}y + \frac{1}{\sqrt{\Omega}}\xi_x(t)\\
\frac{{\rm d}y}{{\rm d}t}&=& m_{21}x + m_{22}y + \frac{1}{\sqrt{\Omega}}\xi_y(t)
\end{eqnarray}
where $\Omega$ scales as the system size, $m_{ij}$ are the elements of the matrix $\mathbf{M}$ given in Eq. (\ref{M}) and $\xi_x(t)$ and $\xi_y(t)$ are zero-mean Langevin forces. Due to the linearization of the Langevin equations, we have $\langle X \rangle=X_0^0$ and $\langle Y \rangle=Y_0^0$
and  $\langle x \rangle=\langle y \rangle=0$.
The variances and covariance of the Langevin forces evaluated at the steady state are given by
\begin{eqnarray}
\langle \xi_x(t) \xi_x(t') \rangle &=& F_{xx} \delta(t-t')\\
\langle \xi_x(t) \xi_y(t') \rangle &=& F_{xy} \delta(t-t')\\
\langle \xi_y(t) \xi_y(t') \rangle &=& F_{yy} \delta(t-t')
\end{eqnarray}
with \cite{gillange}
\begin{eqnarray}
F_{xx}&=&k_1X_0^0+k_2(X_0^0)^2Y_0^0  \\ 
F_{xy}&=&-k_2(X_0^0)^2Y_0^0  \\
F_{yy}&=&k_2(X_0^0)^2Y_0^0+k_3Y_0^0+k_{-3}
\end{eqnarray}
Using the change of basis matrix $\mathbf{P^0}$ given in Eq. (\ref{P}), we determine the scaled variances and covariance 
\begin{eqnarray}   
\Omega\langle x^2 \rangle&=&(p_{11}^0)^2 F_{11} +2p_{11}^0p_{12}^0F_{12} + (p_{12}^0)^2 F_{22}\\
\Omega\langle xy \rangle&=&p_{11}^0p_{21}^0F_{11} +(p_{11}^0p_{22}^0+p_{12}^0p_{21}^0)F_{12} + p_{12}^0p_{22}^0F_{22}\\
\Omega\langle y^2 \rangle&=&(p_{21}^0)^2 F_{11} +2p_{21}^0p_{22}^0F_{12} + (p_{22}^0)^2 F_{22}
\end{eqnarray}
with
\begin{equation}
F_{ij}=\frac{q_{i1}q_{j1}F_{xx} +(q_{i1}q_{j2}+q_{i2}q_{j1})F_{xy} + q_{i2}q_{j2}F_{yy}}{-\lambda_i-\lambda_j}
\end{equation}
where $i,j=1,2$ and the $q_{ij}$'s are the elements of the inverse matrix of $\mathbf{P}$. 

The chemical master equation for the two-variable model is \cite{nicolis,gardiner}
\begin{eqnarray}
\label{me}
\frac{\partial P}{\partial t}&=&k_1\big[(N_X+1)P(\{N_X+1\})-N_XP\big] \nonumber\\
                             &+&\frac{k_2}{\Omega^2}\big[(N_X-1)(N_X-2)(N_Y+1)P(\{N_X-1,N_Y+1\})-N_X(N_X-1)N_YP\big] \nonumber\\
                             &+&k_3\big[(N_Y+1)P(\{N_Y+1\})-N_YP\big]+k_{-3}\Omega\big[P(\{N_Y-1\})-P\big]
\end{eqnarray}
for $N_X$ and $N_Y$ particles of species X and Y at time $t$ in a system of size $\Omega$. Only the dependence of the probability $P$ on the number of particles X and Y
differing from $N_X$ and $N_Y$, respectively, is explicitly indicated.
The kinetic Monte Carlo algorithm developed by Gillespie~\cite{gillespie} is used to directly simulate the reaction processes and numerically solve the master equation
for a system prepared in the steady state with $N_X(t=0)=\Omega X_0^0$ and $N_Y(t=0)=\Omega Y_0^0$.
The scaled variances $\Omega \langle (N_X-\langle N_X \rangle )^2 \rangle$ and $\Omega\langle (N_Y-\langle N_Y \rangle )^2 \rangle$ and the covariance $\Omega\langle (N_X-\langle N_X \rangle)(N_Y-\langle N_Y \rangle) \rangle$ of the fluctuations are computed using a statistics over $10^4$ realizations.\\

The master equations associated with the three-variable models are given in Appendix C.

\begin{figure}
\centering
\subfigure{\includegraphics[scale=0.4]{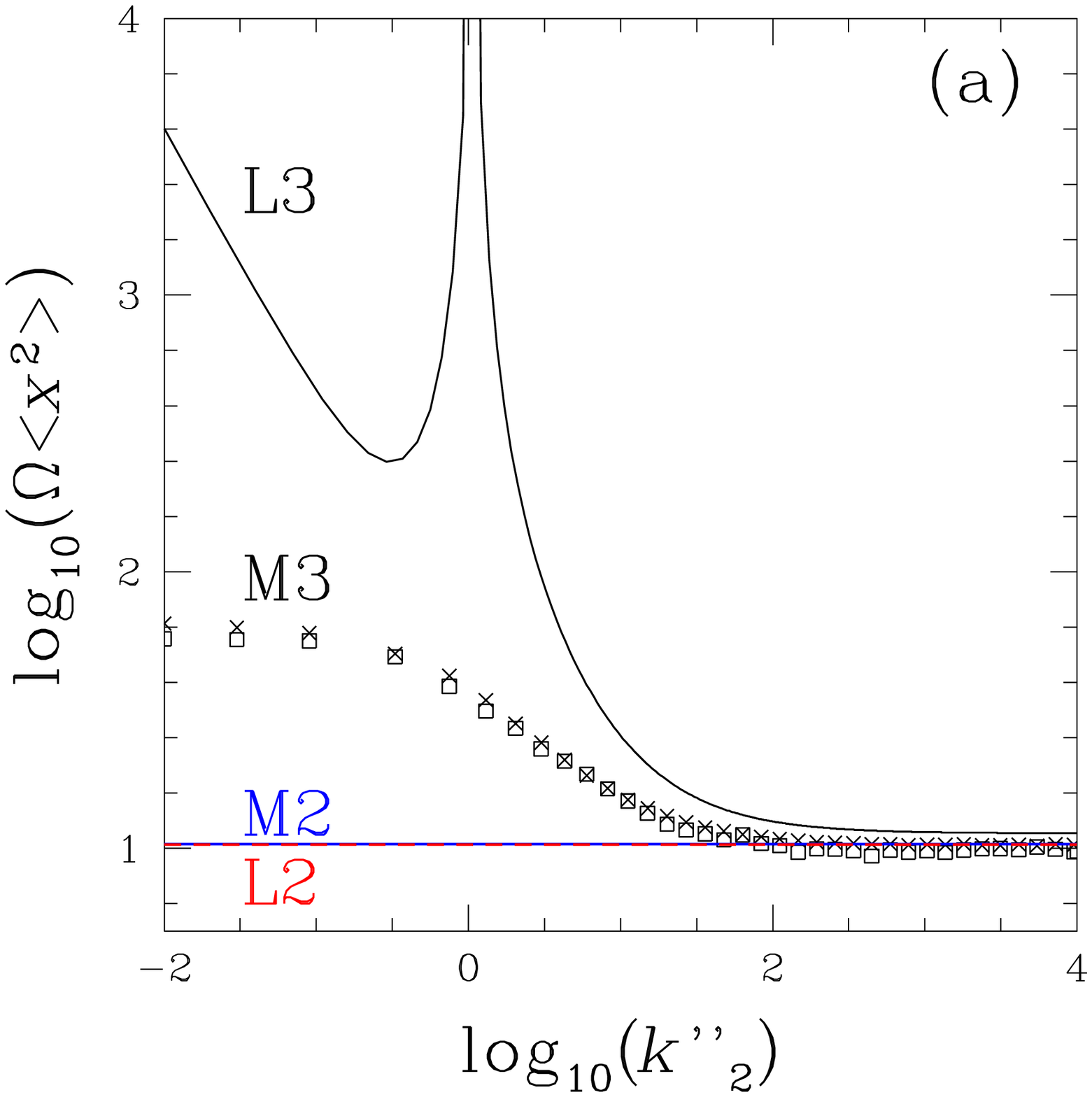}}
\subfigure{\includegraphics[scale=0.4]{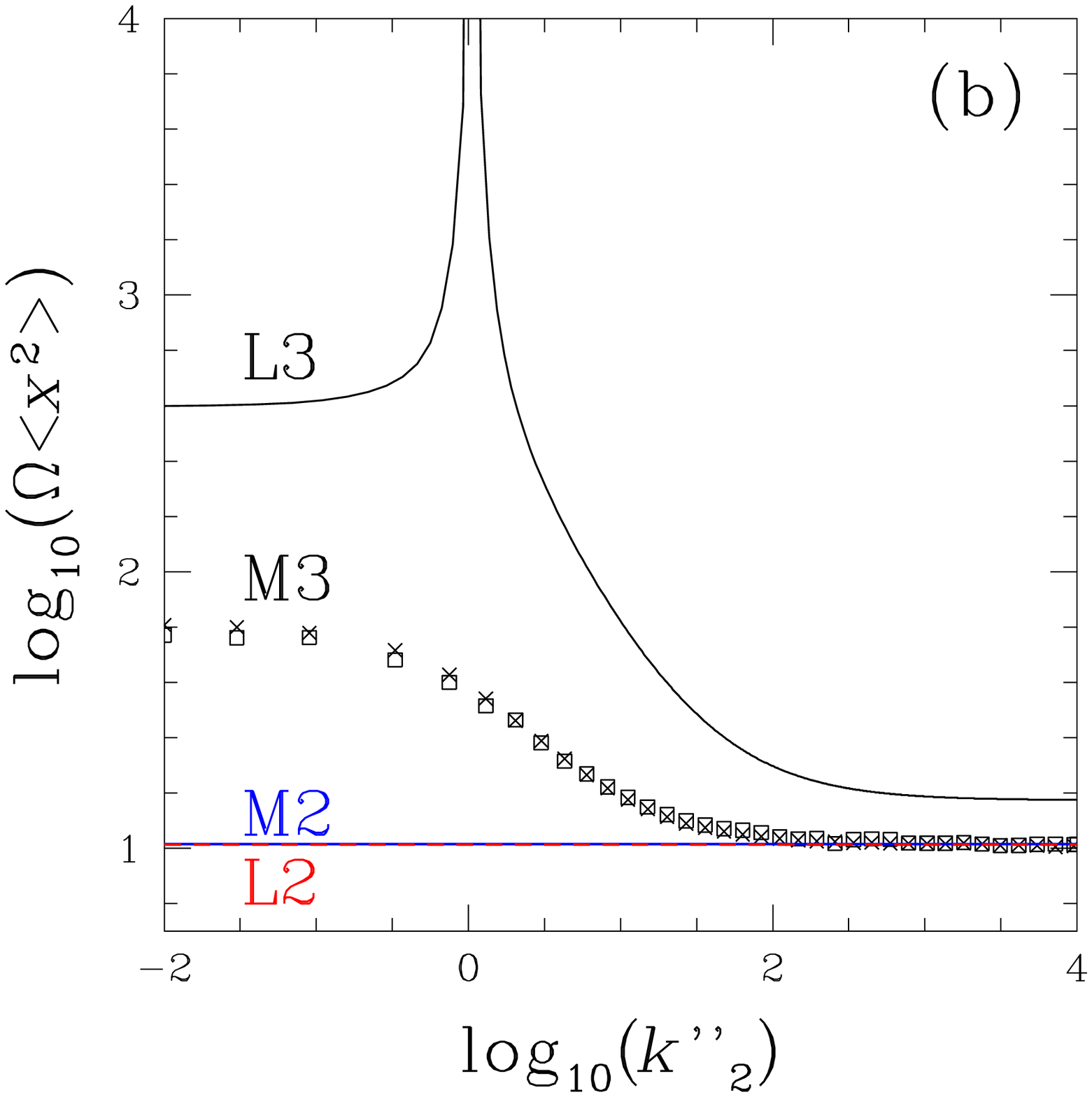}}
\caption{Three-variable model $A$: Logarithm of the scaled variance associated with species X versus $\log_{10}(k''_2)$ for the Langevin approach 
($\log_{10}(\Omega\langle x^2 \rangle)$, black solid line) and the master equation 
($\log_{10}(\Omega \langle (N_X-\langle N_X \rangle )^2 \rangle$) for $\Omega=3$ ($\square$) and $\Omega=1000$ ($\times$). Two-variable model: The red dashed line gives the results of the Langevin approach and the blue solid line, the results of the master equation for $\Omega=1000$. The other parameter values are given in the caption of Fig. 2. (a) Case $a$ and (b) Case $b$.}
%6
\end{figure}
Figure 6ab displays the variation of the scaled variance of the concentration fluctuations of species X versus rate constant $k''_2$ 
for the three-variable model $A$ in cases a and b. 
The variance is multiplied by $\Omega$, so that the scaled results $\Omega\langle x^2 \rangle$ 
deduced from the Langevin approach (see Eq. (\ref{x2a})) are not sensitive to system size and more easily compared to the corresponding result,
$\Omega\langle (N_X-\langle N_X \rangle )^2 \rangle$, deduced from the master equation.
The results of the three-variable model $A$ are compared to those of the two-variable model for both stochastic approaches. In the following, we refer to the Langevin approach applied to the two- and three-variable models as L2 and L3, respectively. The master equation applied to the two- and three-variable models is referred to as M2 and M3.
Due to the logarithmic scale adopted to represent the large range covered by the L3 results, the L2 and M2 approaches seem to coincide in Fig. 6ab. The results of the master equation applied to the two-variable model depend little on system size $\Omega$ at the scale of the figure and only the results obtained for $\Omega=1000$ are shown. Indeed, the parameter values of the two-variable model are chosen far from any situation associated with large fluctuation effects such as bifurcations. In this kind of situations, the results of the linearized Langevin approach are satisfying and the agreement with the results of the master equation are expected even for quite small system sizes \cite{jcp140,jcp141,physica15}.

No appreciable deviation is observed between the M3 results obtained for different values of system size $\Omega$ in both cases $a$ and $b$.
At figure scale, the M3 results converge to the L2 and M2 result for large $k''_2$ values ($k''_2>10^{2.48}$) and deviate from it as $k''_2$ decreases. 
Typically, for $k''_2 \simeq 1$, the scaled variance deduced from M3 is multiplied by a factor of 3 with respect to the M2 results. 
When compared to M3, the L3 results are found to markedly overestimate the deviation between the results of the three- and two-variable models in the range $k''_2 < 10$. 
The L3 results display a spurious divergence for $k''_2=1$, in relation to the crossing of the two eigenvalues $\lambda_1$ and $\lambda_2$ observed in Fig. 2. 
In spite of the logarithmic scale, a discrepancy between the L3 result and the L2, M2, and M3 results can be observed in the limit of the largest investigated $k''_2$ values, in particular in Fig. 6b. The discrepancy between the L3 and M3 results has two origins. First, the Langevin approach introduces Gaussian noises, which implies the intrinsic truncation of the cumulants of the probability distribution function to the second order. Second, the analytical expressions of the variances are deduced from the linearization of the Langevin equations around the steady state whereas the master equation provides an exact stochastic description at a mesoscopic scale. \\

\begin{figure}
\centering
\subfigure{\includegraphics[scale=0.4]{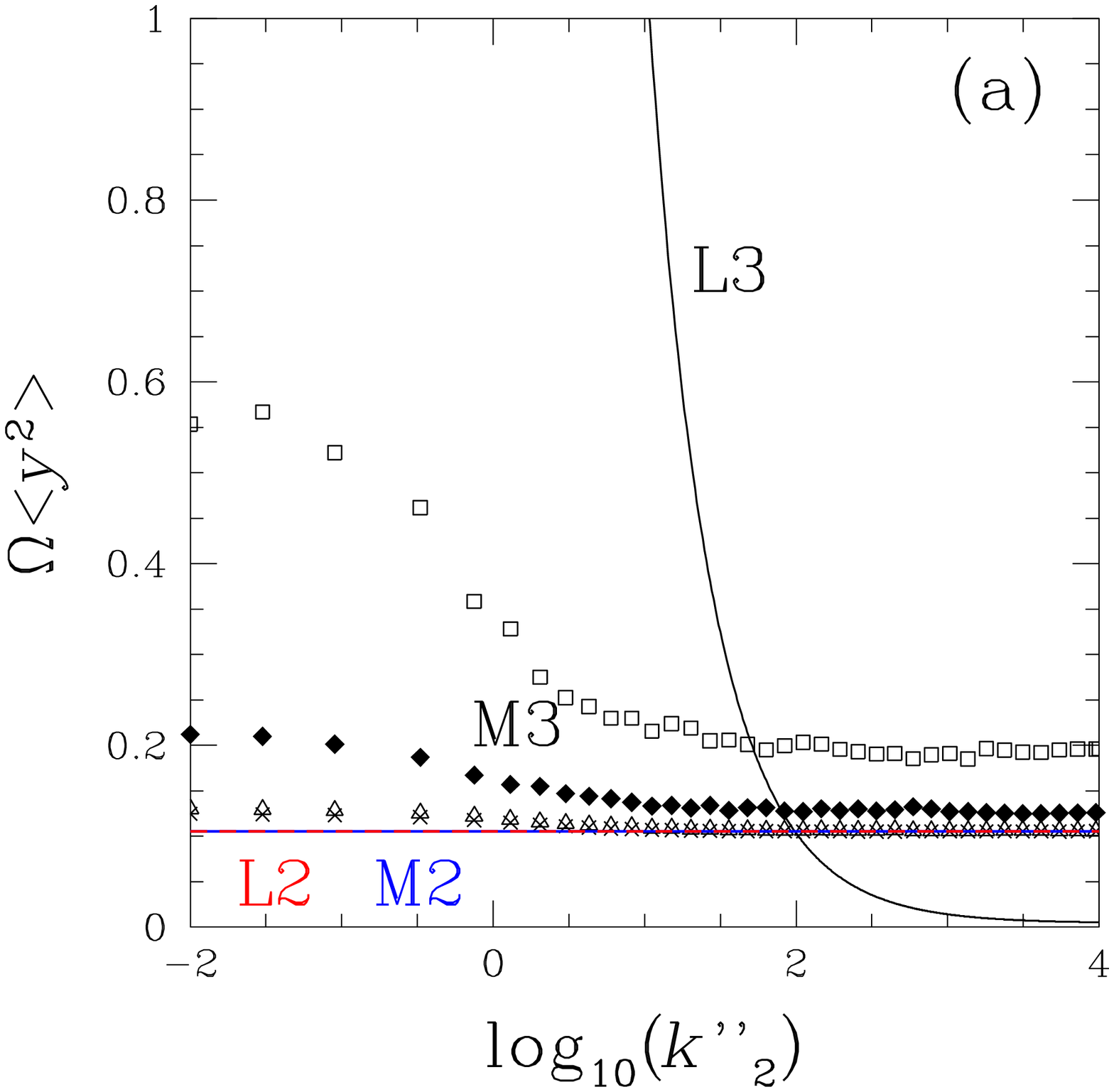}}
\subfigure{\includegraphics[scale=0.4]{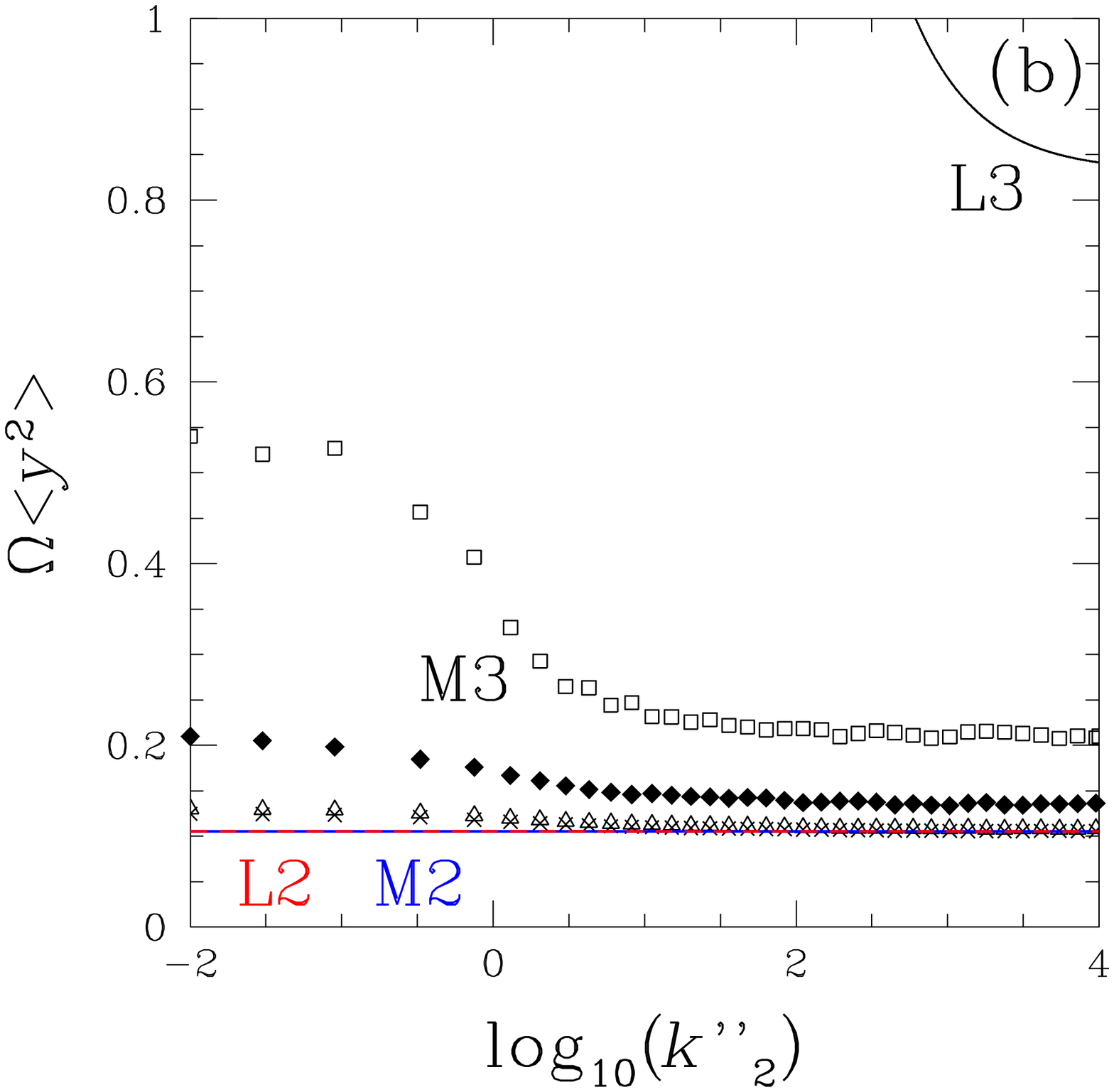}}
\caption{Three-variable model $A$: Scaled variance associated with species Y versus $\log_{10}(k''_2)$ for the Langevin approach 
($\Omega\langle y^2 \rangle$, black solid line) and the master equation ($\Omega \langle (N_Y-\langle N_Y \rangle )^2 \rangle$) 
for $\Omega=3$ ($\square$), $\Omega=10$ ($\blacklozenge$), $\Omega=100$ ($\triangle$), and $\Omega=1000$ ($\times$). Two-variable model: The red dashed line gives the results of the Langevin approach and the blue solid line, the results of the master equation for $\Omega=1000$. The other parameter values are given in the caption of Fig. 2. (a) Case $a$ and (b) Case $b$.}
%7
\end{figure}

The variation of the scaled variance $\Omega \langle y^2 \rangle$ or $\Omega \langle (N_Y-\langle N_Y \rangle )^2 \rangle$ of the concentration fluctuations of species Y with $k''_2$ is given in Fig. 7.  
The two stochastic approaches L2 and M2 for the two-variable model agree. The L3 results are entirely different from both the M3 results and the L2 and M2 results, 
including in the large $k''_2$ limit.
Contrary to the L3 results, the M3 results are similar in cases $a$ and $b$.
The L3 results significantly deviate from the M3 predictions for $k''_2< 10^{1.48}$.
The M3 results converge towards the L2 and M2 results in the limit of large $k''_2$, $(k''_2>100)$, and large system size, ($\Omega \geq 100)$. 
Interestingly, the M3 results obtained for sufficiently small system sizes ($\Omega < 100$) never coincide with the L2 and M2 results: Even in the parameter domain  $k''_2 > 10^{1.48}$, where the quasi-steady-state approximation is valid from the deterministic point of view, the behavior of the variance of species $Y$ in the three-variable model
significantly differs from the behavior obtained in the two-variable model. 

\begin{figure}
\centering
\subfigure{\includegraphics[scale=0.4]{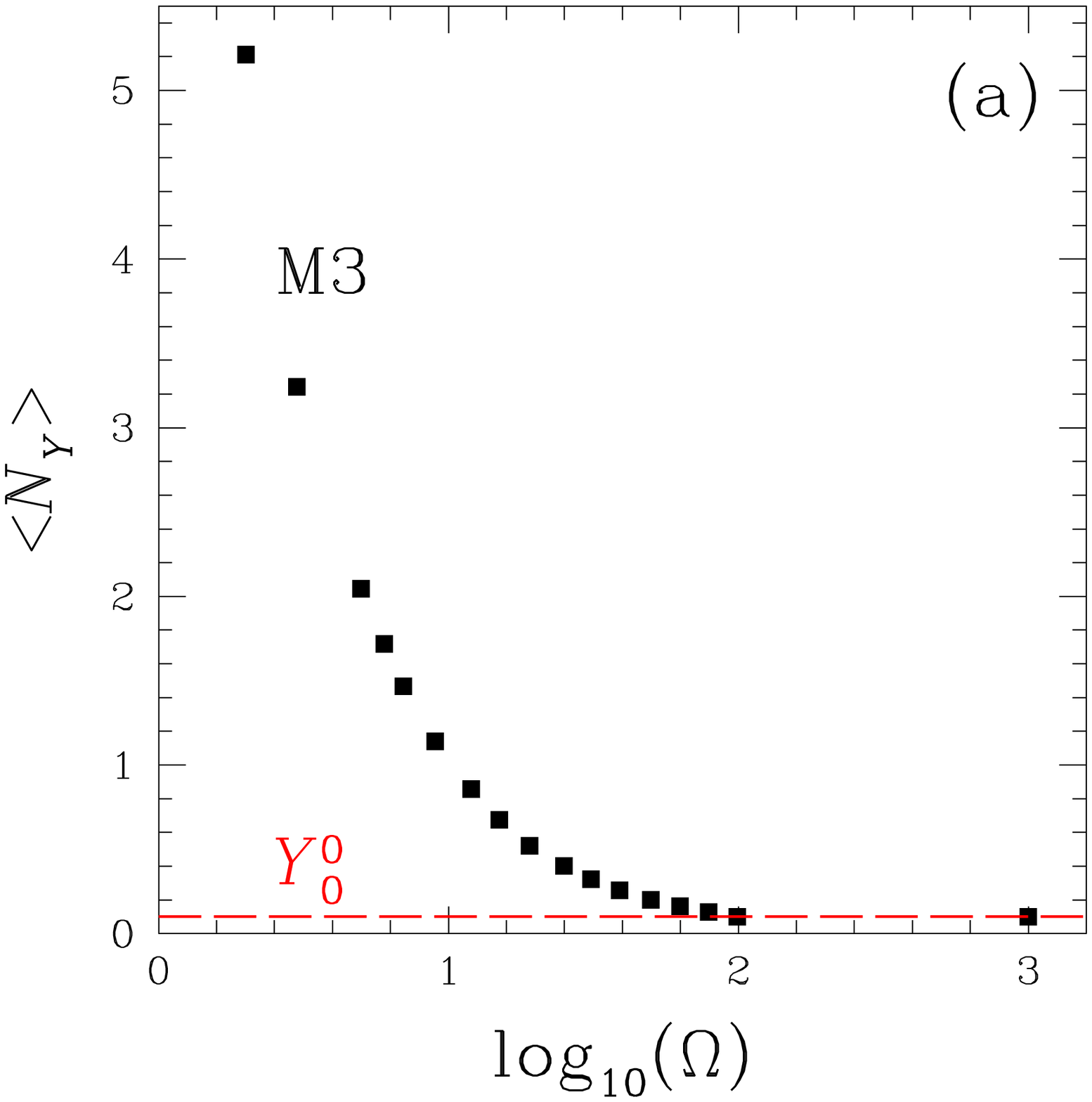}}
\subfigure{\includegraphics[scale=0.4]{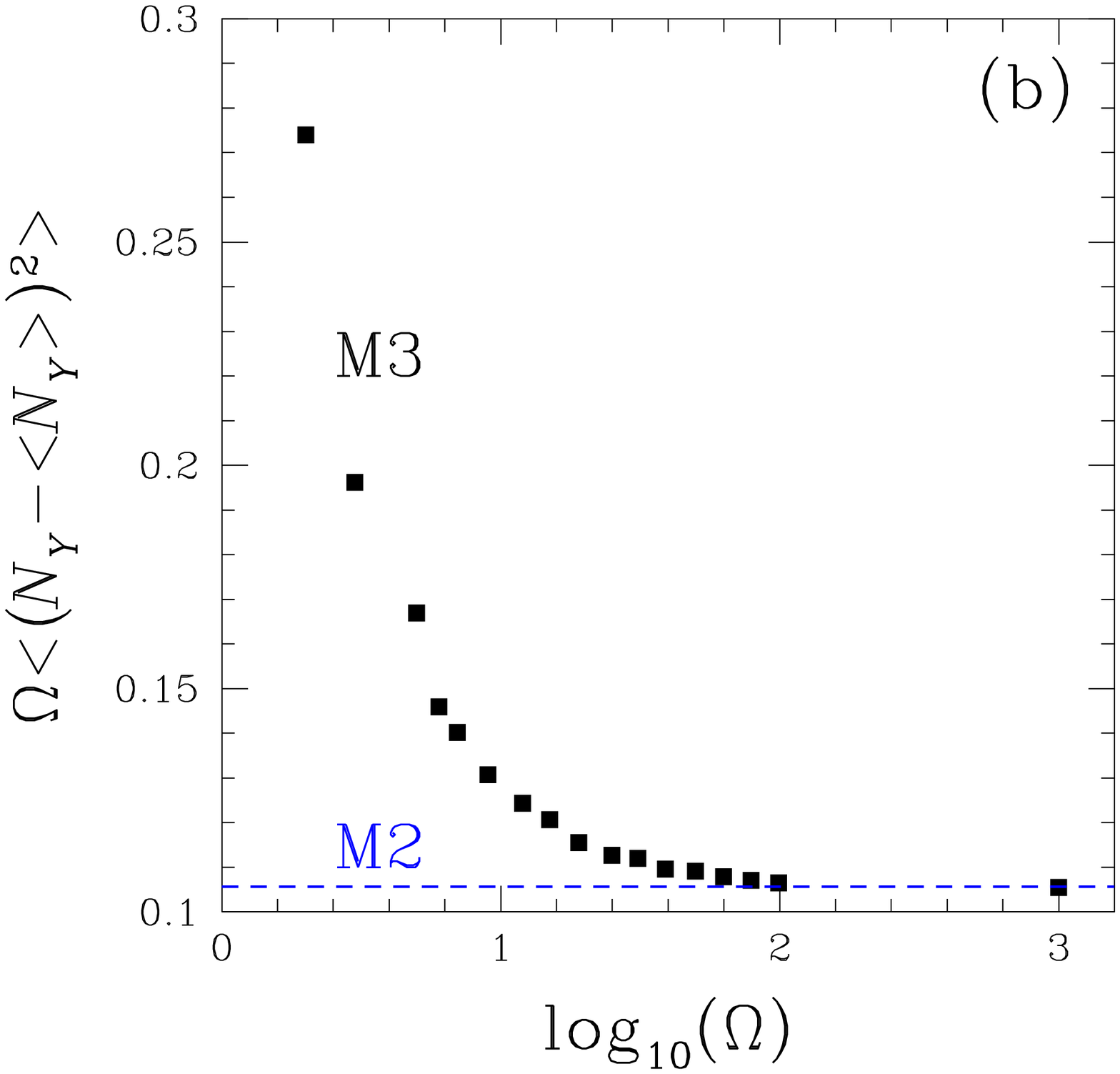}}
\caption{Three-variable model $A$ in case $a$ for $k''_2=10^4$: (a) Mean value $\langle N_Y \rangle$ deduced from the master equation ($\blacksquare$) and deterministic prediction $Y_0^0$ (red long-dashed line) versus $\log_{10}(\Omega)$.
(b) Scaled variance $\Omega\langle (N_Y-\langle N_Y \rangle)^2 \rangle$ deduced from the master equation for the three-variable model $A$ ($\blacksquare$) and the two-variable model (blue short-dashed line) versus $\log_{10}(\Omega)$. The other parameter values are given in the caption of Fig. 2a.}
%8
\end{figure}

The effect is more marked in the case of species Y than species X, due to the smaller value of the steady concentration $Y_0^0$: 
A concentration being a positive variable, the fluctuations around the mean value become asymmetrical when they reach the order of $Y_0^0$, 
i.e. for a sufficiently small system size $\Omega$. As shown in Fig. 8, this phenomenon leads to the increase of the scaled variance 
$\Omega \langle (N_Y-\langle N_Y \rangle)^2 \rangle$ as $\Omega$ decreases. In addition, the mean value $\langle N_Y \rangle$ is shifted 
from the deterministic steady state $Y_0^0$. \\

\begin{figure}
\centering
\subfigure{\includegraphics[scale=0.4]{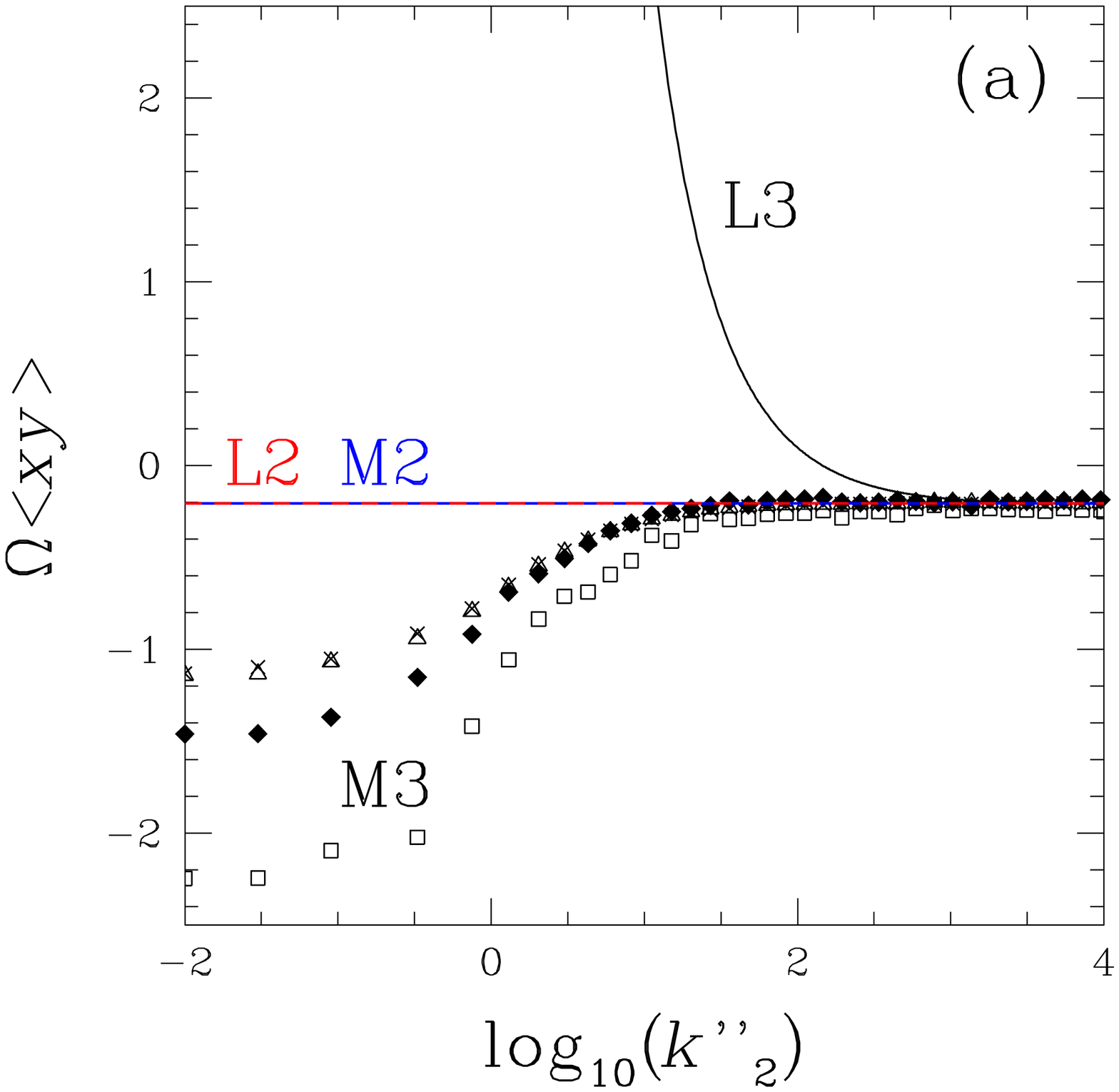}}
\subfigure{\includegraphics[scale=0.4]{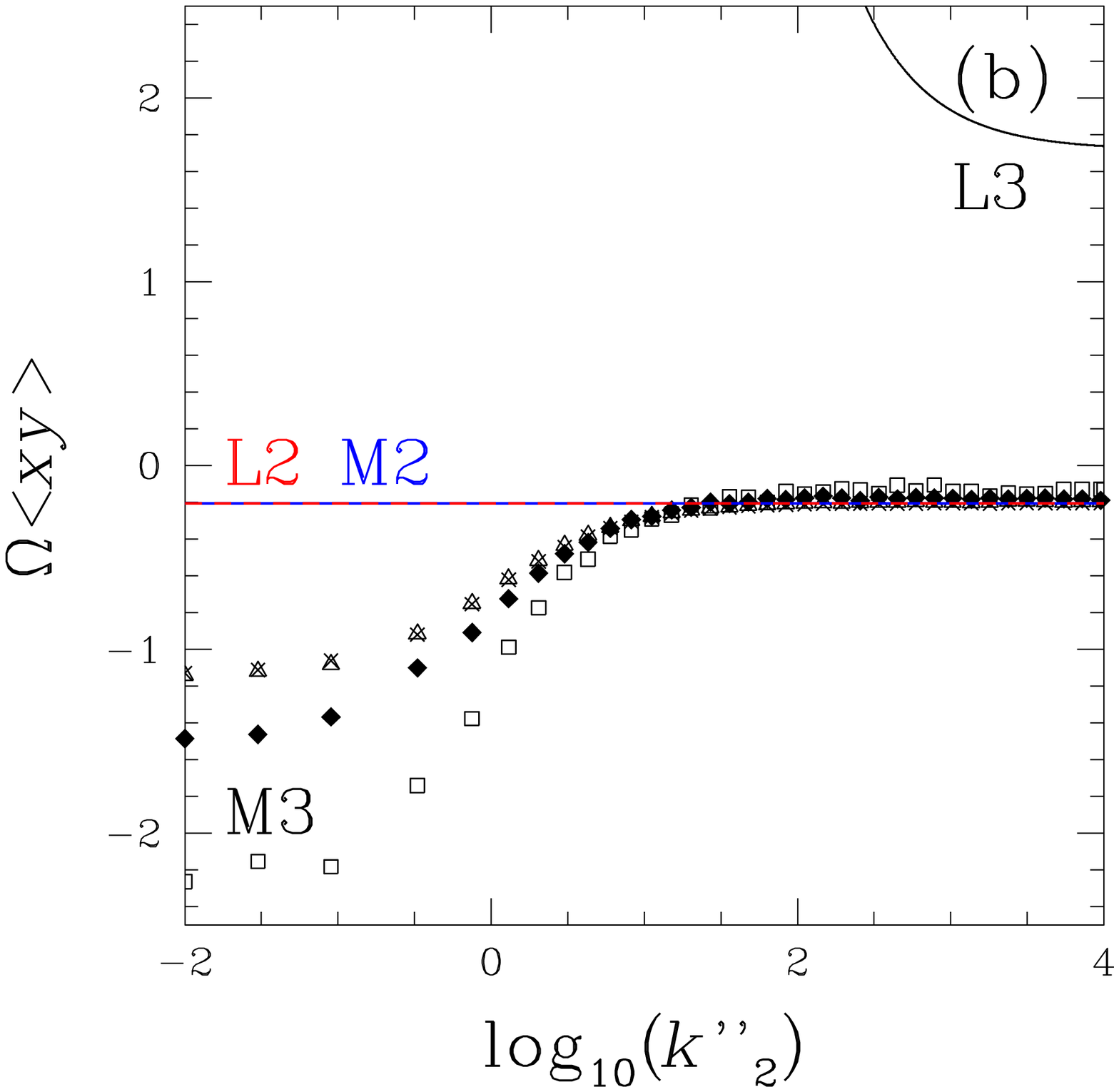}}
\caption{Three-variable model $A$: Same caption as in Fig. 7 for the scaled covariance $\Omega \langle xy \rangle$ (Langevin approach) or $\Omega \langle (N_X-\langle N_X \rangle)(N_Y-\langle N_Y \rangle)\rangle$ (master equation). (a) Case $a$ and (b) Case $b$.
}
%9
\end{figure}

The scaled covariance $\Omega \langle xy \rangle$ or 
$\Omega\langle (N_X-\langle N_X \rangle)(N_Y-\langle N_Y \rangle) \rangle$ associated with species X and Y is represented in Fig. 9. In case $a$, 
the results of the two stochastic approaches for the three-variable model converge toward the results obtained for the two-variable model 
in the limit of large $k''_2$. However, as $k''_2$ becomes smaller than $1000$, the L3 results increase and even become positive whereas the M3 results decrease and begin to appreciably depart from the M2 results only as $k''_2$ becomes smaller than $10$.
In case $b$, the L3 results are always far from the L2 results. 
As shown in Fig. 2, the convergence of the eigenvalue $\lambda_2$ of the three-variable model $A$ toward the corresponding eigenvalue $\lambda_2^0$ of the two-variable model as $k''_2$ increases is slower in case $b$ than in case $a$. Similarly, the increase of the absolute element $\mid q_{33}\mid$ as $k''_2$ increases is slower in case $b$ than in case $a$, as observed in Fig. 3. This could explain that, for large values of $k''_2$, the Langevin approach L3 is more different from L2 in case $b$ than in case $a$ in Figs. 6, 7, and 9. The master equation approach is less sensitive to this feature. Only a small increase of the M3 results with respect to the M2 results is detectable in Fig. 9b in the limit of large $k''_2$ for a small system size $\Omega=3$, whereas the corresponding L3 results are much larger than the L2 results. These observations suggest that the nonlinearities ignored in the linearized Langevin approach mitigate the effect of the discrepancies between the eigenvalues of the two- and three-variable models.
The sign of the covariance can be easily deduced for the two-variable model, in which the second reaction given in Eq. (\ref{reac2II}) forms species X and consumes species Y: The covariance is negative.
The situation is different for the three-variable model $A$, in which both species X and Y are consumed (formed, resp.) by the forward (backward, resp.) reaction 
given in Eq. (\ref{reac2IIIa}) whereas only X is formed by the third reaction given in Eq. (\ref{reac3IIIa}). According to Fig. 9, the master equation approach M3 captures the sign of the covariance even for small values of $k''_2$ for which the elimination of $Z$ is not valid whereas L3 fails.\\

Figures 10-12 give the variances and covariance of the fluctuations of the slow variables $X$ and $Y$ 
for the three-variable model $B$. Due to the similarity of the behaviors observed in cases $a$ and $b$, only case $a$ 
is represented. The results of the stochastic approaches are not given in the parameter domain $k''_2<0.1$ in which the steady state is unstable, as shown in Fig. 4a.  
\begin{figure}
\centering
\includegraphics[scale=0.8]{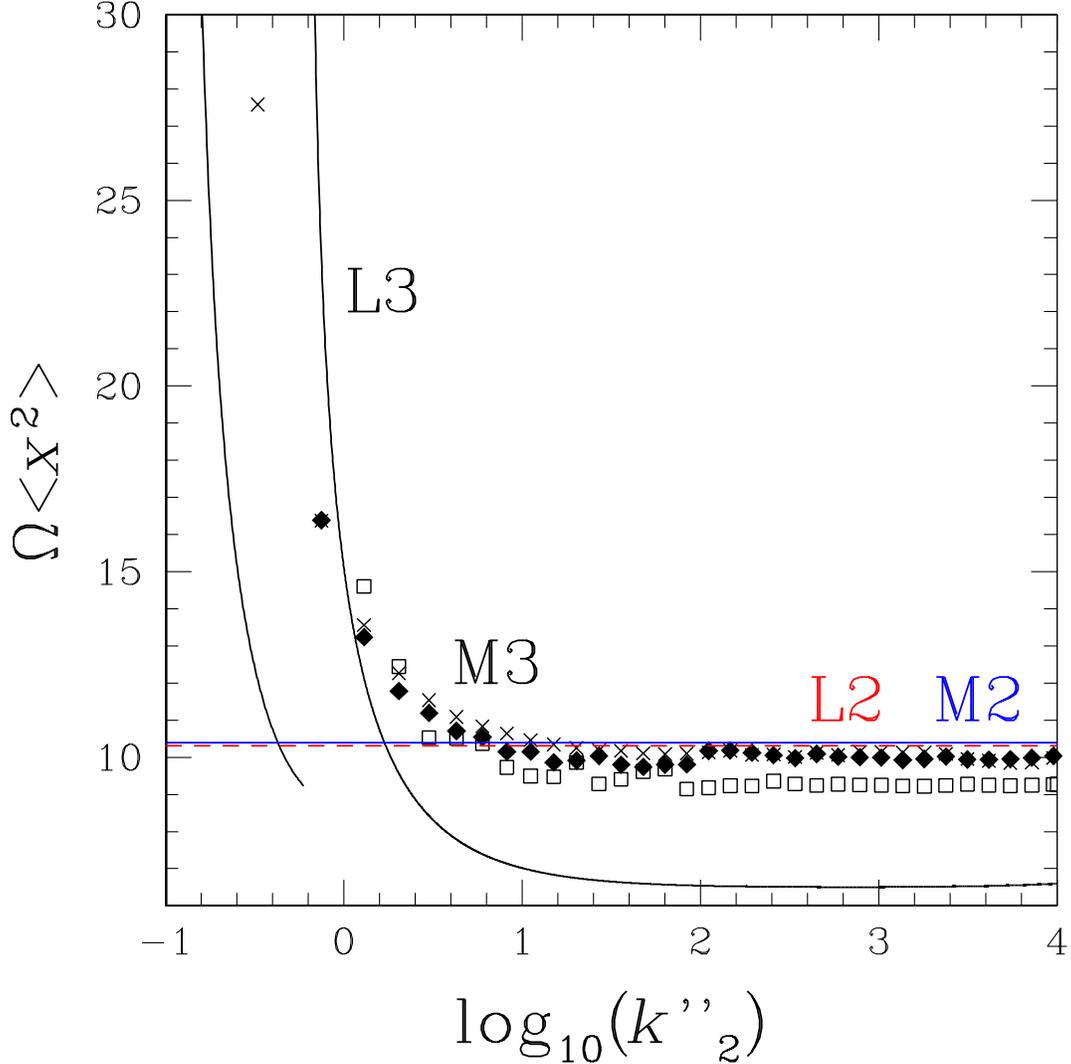}
\caption{Three-variable model $B$ in case $a$: Scaled variance associated with species X versus $\log_{10}(k''_2)$ for the Langevin approach 
($\Omega\langle x^2 \rangle$ , black solid line) and the master equation ($\Omega \langle (N_X-\langle N_X \rangle )^2 \rangle$) 
for $\Omega=3$ ($\square$), $\Omega=10$ ($\blacklozenge$), and $\Omega=1000$ ($\times$). Two-variable model: The red dashed line gives the results of the Langevin approach and the blue solid line, the results of the master equation for $\Omega=1000$. Case $a$: $k'_{-2}=1$ with $k'_2=k_2(X_0^0+k'_{-2}/k''_2)$.
The other parameter values are given in the caption of Fig. 2.}
%10
\end{figure}

According to Fig. 10, the scaled variance $\Omega \langle x^2 \rangle$ deduced from L2 is slightly smaller than the result deduced from M2. 
The small gap between L2 and M2 results could not be seen in Fig. 6 due to the adopted logarithmic scale. 
The L3 results of the Langevin approach applied to the three-variable model completely differ from the L2 predictions in the entire range of $k''_2$: 
The L3 limit at large $k''_2$ underestimates the L2 value by a factor of $1.6$ and the spurious divergence observed for $k''_2=0.6$ due to the coalescence of the eigenvalues $\lambda_1$ and $\lambda_2$ (see Fig. 4) induces a rapid increase of $\Omega \langle x^2 \rangle$ in the range $10^{-0.22}<k''_2<100$.
The M3 results for the scaled variance $\Omega \langle (N_X-\langle N_X \rangle )^2 \rangle$ do not converge 
toward the M2 results even in the limit of large $k''_2$ and $\Omega$. The fact that, for the same parameter values, 
$Z_0$ is smaller in the case of model $B$ than model $A$ (see Fig. 1) is not sufficient to ensure the matching between 
the stochastic predictions of the two-variable model and three-variable model $B$.
Indeed, the nonlinearities observed in the rate laws (see Eqs. (\ref{eqXIIIbred}) and (\ref{eqYIIIbred})) after applying the quasi-steady-state approximation to model $B$ are different 
from the nonlinearities present in the original two-variable model. The nonconvergence of the M3 results toward the M2 results at large $k''_2$ and $\Omega$ 
in the case of model $B$ illustrates the decisive role played by the nonlinearities of the deterministic dynamics in fluctuation properties.

\begin{figure}
\centering
\includegraphics[scale=0.8]{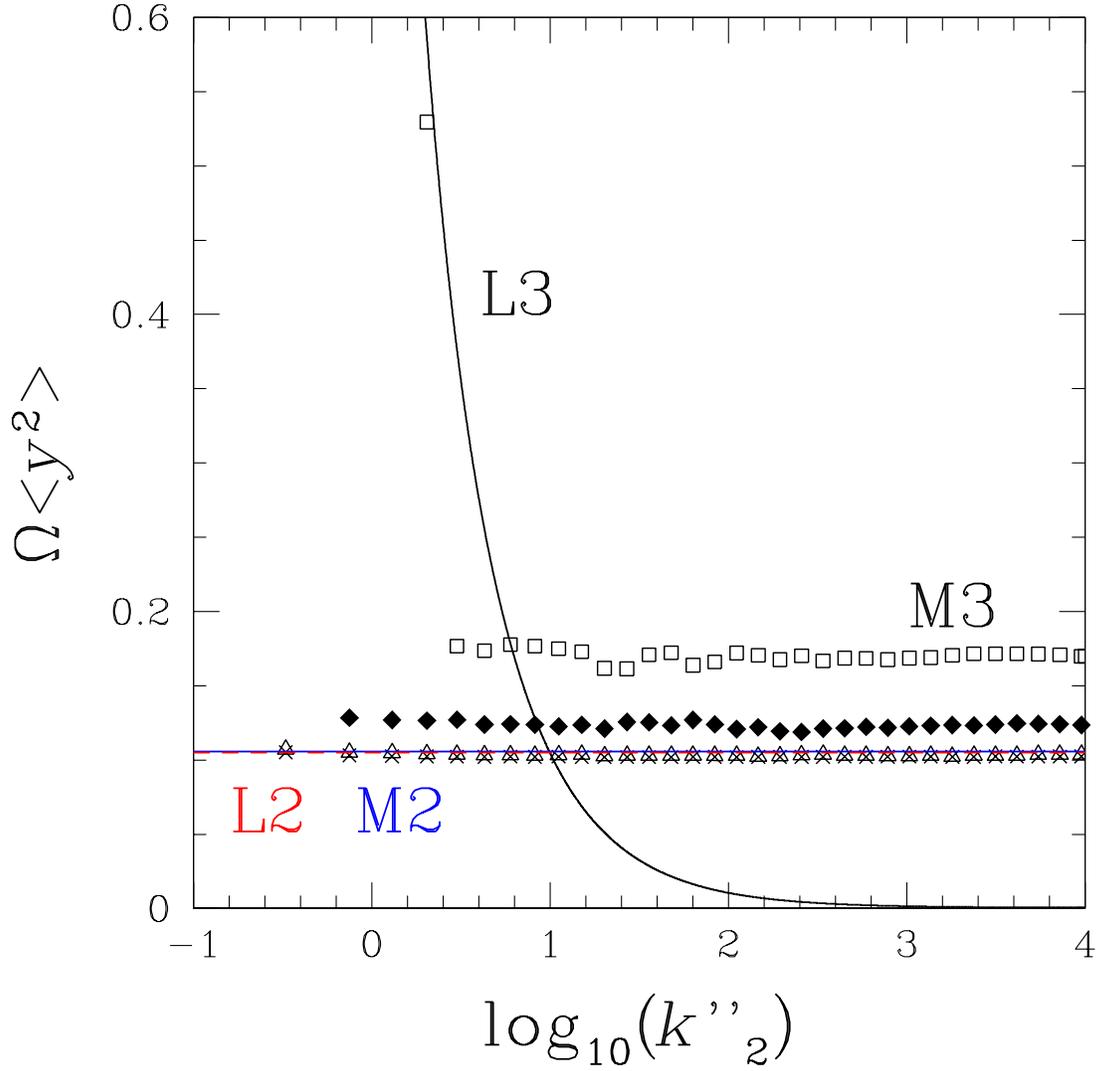}
\caption{Three-variable model $B$: Same caption as in Fig. 10 for the scaled variance $\Omega \langle y^2 \rangle$ (Langevin approach) or $\Omega \langle (N_Y-\langle N_Y \rangle)^2\rangle$ (master equation) associated with species Y.}
%11
\end{figure}
The sensitivity of the M3 results to system size for species X in Fig. 10 is small, compared to the results observed in Fig. 11 for species Y. 
As already mentioned, the sensitivity of $\Omega \langle (N_Y-\langle N_Y \rangle )^2 \rangle$ to $\Omega$ is related to the small value of $Y_0^0$ 
which induces asymmetrical fluctuations around the mean value for sufficiently small system sizes.
The results given in Fig. 11 for model $B$ are qualitatively the same as those given in Fig. 7 for model $A$.\\

\begin{figure}
\centering
\includegraphics[scale=0.8]{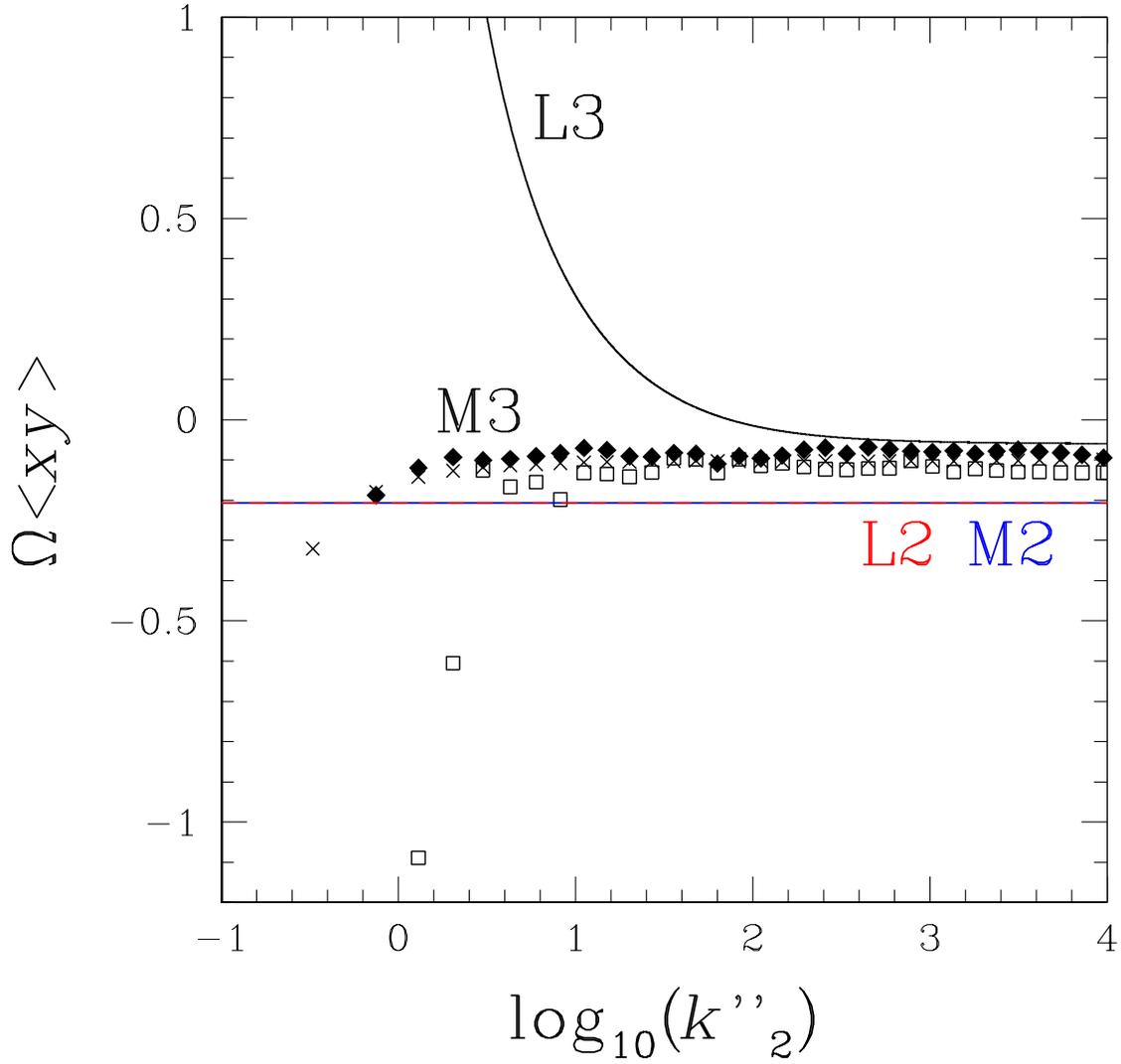}
\caption{Three-variable model $B$: Same caption as in Fig. 10 for the scaled covariance $\Omega \langle xy \rangle$ (Langevin approach) or $\Omega \langle (N_X-\langle N_X \rangle)(N_Y-\langle N_Y \rangle)\rangle$ (master equation).}
%12
\end{figure}
The results obtained for the covariance in the case of model $B$ are given in Fig. 12. Contrary to model $A$ (see Fig. 9) for which the results of the master equation 
converge toward the L2 and M2 limit at large $k''_2$, the M3 results obtained for model $B$ remain larger than the L2 and M2 limit, even for a large system size $\Omega =1000$. The covariance deduced from the M3 approach decreases as $k''_2$ tends to the critical value $k''_2=0.1$ below which the steady state becomes unstable.

Although the three-variable models $A$ and $B$ converge to the same deterministic two-variable model, at least for the steady properties and linear
dynamics, the variances and covariance of the slow variables in models $A$ and $B$ have different behaviors.
Hence, the proper description of the fluctuations in models $A$ and $B$ by the two-variable model deduced from the steady-state-approximation is not ensured in systems of small size. Sizes of the order of $\Omega=1000$ have to be reached for the results of the master equation applied to the three-variable models to converge toward the results of the two-variable model.

\section{Discussion and conclusion}
In this paper, we introduce a fast species in a chemical mechanism and show that it perturbs the fluctuations of the slow species even in the domain of validity of the quasi-steady-state approximation, in which the deterministic dynamics is correctly predicted by the reduced mechanism in the linear domain. \\

The study provides an opportunity to revisit the quasi-steady-state approximation. By completely neglecting fast relaxation and restricting the dynamics to the slow manifold, the approximation can be considered as a zeroth-order perturbation method, which masks the small parameter and makes the definition of the validity domain less obvious.

The existence of a fast variable is sufficient to reduce dynamics but, in the general case, the slow variables are non trivial functions of the entire set of the initial variables, e.g. a linear combination of all concentrations in the framework of a linear analysis.
The quasi-steady-state approximation eliminates a chemical species and consequently requires that the fast variable is close to an actual concentration. 
In the linear domain, characteristic relaxation times are assigned to the evolution along eigendirections and the notion of fast or slow variables to coordinates in the eigenbasis. 
We express the conditions of validity of the approximation using the eigenvalues and appropriate elements of the change of basis matrix. The parallelism with the Born-Oppenheimer approximation in quantum chemistry is drawn, in particular in the case of avoided crossing between two eigenvalues. \\
The approximation is applied to two three-variable models $A$ and $B$ and the possible reduction to a same two-variable model is discussed. Conditions on the parameters of the three-variable models are made explicit for the two- and three-variable models to have the same steady state of interest.
Model $B$ offers a simple example showing that small concentrations for a slowly formed and rapidly consumed species cannot be considered as a valid condition of elimination. The intuitive belief used to eliminate a very reactive intermediate from a mechanism coincides with the dynamical criteria made precise in section 2 in the case of intrinsically linear dynamics but not in general.
In the case of model $A$, the steady value $Z_0 \simeq 2.5$ of the eliminated variable $Z$ that is reached for $k''_2=10^4$ is already reached for $k''_2=10^{0.3}$ in the case of model $B$. According to the master equation and for a given value of $Z_0$, the deviations of the variances and covariance 
in the three-variable models to the corresponding quantities in the two-variable model are smaller for model $A$ than model $B$.
More precisely, for a large system size, $\Omega=1000$, the M3 value of the variance $\Omega \langle (N_X-\langle N_X \rangle)^2 \rangle$ in model $A$ for $k''_2=10^4$ underestimates the M2 value by less than $1\%$ whereas the corresponding result associated with the same $Z_0$ value in model $B$, i.e. for $k''_2=10^{0.3}$, overestimates the M2 value by $18\%$.
Similarly, the M3 result for the covariance $\Omega \langle (N_X-\langle N_X \rangle) (N_Y-\langle N_Y \rangle)\rangle$ in model $A$ for $k''_2=10^4$ overestimates the M2 value by only $0.6\%$ whereas the corresponding result in the case of model $B$ for $k''_2=10^{0.3}$ overestimates the M2 value by $38\%$. 
The larger discrepancies between M3 and M2 results in the case of model $B$ for the same value of $Z_0$ are related to the specific nonlinearities introduced in the macroscopic rate equations when eliminating the fast variable $Z$. These nonpolynomial nonlinearities differ from the nonlinearities of the two-variable model and the interplay between the nonlinearities of the deterministic dynamics and the fluctuations is known to be complex and model specific \cite{jcp140,jcp141,physica15}. The two-variable mechanism does not correctly account for the nonlinearities of the reduced dynamics of the three-variable model $B$ and can only claim to model the linearized properties of the three-variable model around the steady state.
The reduction of a mechanism often leads to such nonpolynomial nonlinearities, as for example in the case of the reduced Michaelis-Menten scheme \cite{michaelis}. Our results show that the conclusions that could be deduced from a stochastic analysis relying on the reduced Michaelis-Menten model may differ from the direct analysis of the complete scheme. \\
 
In the parameter range in which the reduction of the deterministic dynamics is valid in the linear domain and for both three-variable models $A$ and $B$, the variances and covariance of the fluctuations of the slow concentrations do not always coincide with the 
corresponding quantities obtained for the two-variable model. The deficiencies of the linearized Langevin approach 
in capturing the properties of the internal fluctuations in a nonlinear chemical system and the necessary resort to the master equation 
are pointed out. Even the sign of the covariance of the fluctuations is not always correctly predicted by the Langevin equations. 
The small value $Y_0^0$ of the steady concentration of species Y inducing asymmetrical fluctuations, 
the variance of the $Y$-fluctuations deduced from both three-variable models significantly depends on system size and 
differs from the prediction of the two-variable model for sufficiently small system sizes leading to fluctuation amplitudes larger than $Y_0^0$. 
The variance of the fluctuations around the large steady concentration $X_0^0$ deduced from the three-variable model $B$ is not very sensitive to system size but does not converge toward the prediction of the two-variable model. 
We already pointed out the differences between the nonlinearities of the reduced dynamics obtained for model $B$ and those of the two-variable model of reference. The coupling between the fluctuations and the nonlinearities of deterministic dynamics makes the use of the quasi-steady-state approximation delicate when the studied system requires a good control.
The predictions of a reduced mechanism must be considered with special care when modeling pattern formation in biology, preventing hazards in explosive phenomena, or dealing with small systems in which variances of fluctuations are detected as in fluorescence correlation spectroscopy (FCS).

\begin{appendices}
\section{Eigenvalues and eigenbasis of the three-variable models}
The eigenvalues of the three-variable models are solution to a cubic polynomial, determined through the 
Cardano's method.
In the case of the three-variable model $A$, the stability matrix ${\mathbf M}$ is
\begin{eqnarray}
\label{Ma}
{\mathbf M}=
\begin{pmatrix}
m_{11}=-(4k'_2X_0^0Y_0^0+k_1) & m_{12}=-2k'_2(X_0^0)^2      & m_{13}=2k'_{-2}+3k''_2 \\
m_{21}=-2k'_2X_0^0Y_0^0       & m_{22}=-(k'_2(X_0^0)^2+k_3) & m_{23}=k'_{-2} \\
m_{31}=2k'_2X_0^0Y_0^0        & m_{32}=k'_2(X_0^0)^2        & m_{33}=-(k'_{-2}+k''_2) 
\end{pmatrix}
\end{eqnarray}
with $X_0^0$, $Y_0^0$ given in Eqs. (\ref{X0II}) and (\ref{Y0II}).

For the three-variable model $B$, the linear stability operator ${\mathbf M}$ is
\begin{eqnarray}
\label{Mb}
{\mathbf M}=
\begin{pmatrix}
m_{11}=-k'_2Y_0^0-k_1+2k''_{2}Z_0 & m_{12}=-k'_2X_0^0       & m_{13}=k'_{-2}+2k''_2X_0^0 \\
m_{21}=-k'_2Y_0^0                 & m_{22}=-(k'_2X_0^0+k_3) & m_{23}=k'_{-2} \\
m_{31}=k'_2Y_0^0-k''_{2}Z_0       & m_{32}=k'_2X_0^0        & m_{33}=-(k'_{-2}+k''_2X_0^0) 
\end{pmatrix}
\end{eqnarray}
where the steady concentrations $X_0^0$, $Y_0^0$ are given in Eqs. (\ref{X0II}) and (\ref{Y0II}) and $Z_0$ is given in Eq. (\ref{Z0IIIb}).

Following Cardano's method, we set for each three-variable model
\begin{eqnarray}
a&=&1, \quad b=-(m_{11}+m_{22}+m_{33}),\\
c&=&-m_{12}m_{21}-m_{13}m_{31}-m_{23}m_{32}+m_{11}m_{22}+m_{11}m_{33}+m_{22}m_{33},\\
d&=&m_{11}(m_{23}m_{32}-m_{22}m_{33})+m_{21}(m_{12}m_{33}-m_{13}m_{32})+m_{31}(m_{13}m_{22}-m_{12}m_{23})\\
e&=&-\frac{b^2}{3a^2}+\frac{c}{a}, \\
f&=&\frac{b}{27a}\left(2\frac{b^2}{a^2}-9\frac{c}{a}\right)+\frac{d}{a}, \quad \delta=-(4e^3+27f^2)
\end{eqnarray}
If $\delta \geq 0$, we set 
\begin{equation}
u^3=\frac{-f+i\sqrt{\delta/27}}{2}, \quad v^3=\frac{-q-i\sqrt{\delta/27}}{2}
\end{equation}
and if $\delta < 0$, we write 
\begin{equation}
u^3=\frac{-f+\sqrt{-\delta/27}}{2}, \quad v^3=\frac{-q-\sqrt{-\delta/27}}{2}
\end{equation}
The eigenvalues of the $3\times3$ matrix ${\mathbf M}$ are given by
\begin{eqnarray}
\label{lam1}
\lambda_1&=&\ell u+\bar{\ell}v-\frac{b}{3a} \\
\label{lam2}
\lambda_2&=&u+v-\frac{b}{3a} \\
\label{lam3}
\lambda_3&=&\bar{\ell}u+\ell v-\frac{b}{3a}
\end{eqnarray}
where $\ell=-1/2+i\sqrt{3}/2$.

The $j^{\rm th}$ eigenvector, i.e., the $j^{\rm th}$ column ${\mathbf P_j}$ for $j=1,2,3$ of the change of basis matrix ${\mathbf P}$ can also be written in a form valid for both three-variable models
\begin{eqnarray}
\label{P3}
{\mathbf P_j}=
\begin{pmatrix}
p_{1j}=1  \\
p_{2j}=\frac{\lambda_j-m_{11}-m_{13}p_{31}}{m_{12}}  \\
p_{3j}=\frac{(\lambda_j-m_{11})(\lambda_j-m_{22})-m_{12}m_{21}}{m_{13}(\lambda_j-m_{22})+m_{12}m_{23}} 
\end{pmatrix}
\end{eqnarray}
where the eigenvalues $\lambda_j$ are given in Eqs. (\ref{lam1}-\ref{lam3}) and the elements $m_{ij}$ are given in Eq. (\ref{Ma}) for model $A$ 
and Eq. (\ref{Mb}) for model $B$.

\section{Variances and covariance of concentration fluctuations in the three-variable models deduced from the Langevin equations} 
Extending the approach given in section 3 for a two-variable system to three-variable systems, 
we obtain the scaled variances and covariance of the deviations $x$ and $y$ to the steady concentrations $X_0^0$ and $Y_0^0$
\begin{eqnarray}
\label{x2a}
\Omega\langle x^2 \rangle &=& (p_{11})^2F_{11}+(p_{12})^2F_{22}+(p_{13})^2F_{33}+2p_{11}p_{12}F_{12}+2p_{11}p_{13}F_{13}+2p_{12}p_{13}F_{23} \\
\label{y2a}
\Omega\langle y^2 \rangle &=& (p_{21})^2F_{11}+(p_{22})^2F_{22}+(p_{23})^2F_{33}+2p_{21}p_{22}F_{12}+2p_{21}p_{23}F_{13}+2p_{22}p_{23}F_{23} \\
\label{xya}
\Omega\langle xy \rangle &=& p_{11}(p_{21}F_{11}+p_{22}F_{12}+p_{23}F_{13})
                      +p_{12}(p_{21}F_{12}+p_{22}F_{22}+p_{23}F_{23}) \nonumber \\
                   & &+p_{13}(p_{21}F_{13}+p_{22}F_{23}+p_{23}F_{33})
\end{eqnarray}
with
\begin{eqnarray}
F_{ij}= \frac{q_{i1}(q_{j1}F_{xx}+q_{j2}F_{xy}+q_{j3}F_{xz})+q_{i2}(q_{j1}F_{xy}+q_{j2}F_{yy}+q_{j3}F_{yz})+q_{i3}(q_{j1}F_{xz}+q_{j2}F_{yz}+q_{j3}F_{zz})}{-(\lambda_i+\lambda_j)}
\end{eqnarray}
for $i,j=1,2,3$ and where $q_{ij}$ are the elements of the inverse matrix of the change of basis matrix $\mathbf{P}$ (see Eq. (\ref{P3}))
and the variances and covariances of the Langevin forces are
\begin{eqnarray}
F_{xx}&=&k_1X_0^0+4k'_2(X_0^0)^2Y_0^0+(4k'_{-2}+9k''_2)Z_0\\
F_{yy}&=&k'_2(X_0^0)^2Y_0^0+k'_{-2}Z_0+k_3Y_0+k_{-3}\\
F_{zz}&=&k'_2(X_0^0)^2Y_0^0+(k'_{-2}+k''_2)Z_0\\
F_{xy}&=&2k'_2(X_0^0)^2Y_0^0+2k'_{-2}Z_0\\
F_{xz}&=&-2k'_2(X_0^0)^2Y_0^0-(2k'_{-2}+3k''_2)Z_0\\
F_{yz}&=&-k'_2(X_0^0)^2Y_0^0-k'_{-2}Z_0
\end{eqnarray}
in the case of model $A$ and
\begin{eqnarray}
F_{xx}&=&k_1X_0^0+k'_2X_0^0Y_0^0+(k'_{-2}+4k''_2X_0^0)Z_0\\
F_{yy}&=&k'_2X_0^0Y_0^0+k'_{-2}Z_0+k_3Y_0^0+k_{-3}\\
F_{zz}&=&k'_2X_0^0Y_0^0+(k'_{-2}+k''_2X_0^0)Z_0\\
F_{xy}&=&k'_2X_0^0Y_0^0+k'_{-2}Z_0\\
F_{xz}&=&-k'_2X_0^0Y_0^0-(k'_{-2}+2k''_2)Z_0\\
F_{yz}&=&-k'_2X_0^0Y_0^0-k'_{-2}Z_0
\end{eqnarray}
in the case of model $B$.

\section{Master equation of the three-variable models}
The master equation associated with the three-variable model $A$ is
\begin{eqnarray}
\label{me3a}
\frac{\partial P}{\partial t}&=&k_1\big[(N_X+1)P(N_X+1)-N_XP\big] \nonumber \\
                             &+&\frac{k'_2}{\Omega^2}\big[(N_X+2)(N_X+1)(N_Y+1)P(N_X+2,N_Y+1,N_Z-1)-N_X(N_X-1)N_YP\big] \nonumber \\
                             &+&k'_{-2}\big[(N_Z+1)P(N_X-2,N_Y-1,N_Z+1)-N_ZP\big] \nonumber \\
                             &+&k''_2\big[(N_Z+1)P(N_X-3,N_Z+1)- N_ZP\big] \nonumber \\
                             &+&k_3\big[(N_Y+1)P(N_Y+1)-N_YP\big]+k_{-3}\Omega\big[P(N_Y-1)-P\big]
\end{eqnarray}

The master equation for model $B$ is
\begin{eqnarray}
\label{me3b}
\frac{\partial P}{\partial t}&=&k_1\big[(N_X+1)P(N_X+1)-N_XP\big]\nonumber\\
                             &+&\frac{k'_2}{\Omega}\big[(N_X+1)(N_Y+1)P(N_X+1,N_Y+1,N_Z-1)-N_XN_YP\big] \nonumber\\
                             &+&k'_{-2}\big[(N_Z+1)P(N_X-1,N_Y-1,N_Z+1)-N_ZP\big]\nonumber\\
                             &+&\frac{k''_2}{\Omega}\big[(N_X-2)(N_Z+1)P(N_X-2,N_Z+1)-N_XN_ZP\big]\nonumber\\
                             &+&k_3\big[(N_Y+1)P(N_Y+1)-N_YP\big]+k_{-3}\Omega\big[P(N_Y-1)-P\big]
\end{eqnarray}
\end{appendices}

\section*{Conflicts of interest}
There are no conflicts to declare.

\section{Acknowledgments}
This publication is part of a project that has received
funding from the European Union’s Horizon 2020 (H2020-EU.1.3.4.) research and innovation program under the Marie
Sklodowska-Curie Actions (MSCA-COFUND ID 711859) and from the Ministry of Science and Higher Education, Poland,
for the implementation of an international cofinanced project.

\end{document}